\documentclass[fleqn,usenatbib]{mnras}

\usepackage{newtxtext,newtxmath}

\usepackage[T1]{fontenc}
\usepackage{ae,aecompl}
\usepackage[normalem]{ulem}
  
\usepackage{float}

\usepackage{graphicx}	
\usepackage{amsmath}	
\usepackage[dvipsnames]{xcolor}
\usepackage[most]{tcolorbox}

\newcommand{\hl}[1]{#1}

\hypersetup{draft}

\newcommand{\Mpc}{{\textrm{ Mpc}}}
\newcommand{\hMpc}{{\textrm{ $h^{-1}$Mpc}}}
\newcommand{\kms}{{\textrm{ km\,s$^{-1}$}}}

\newcommand{\hMsun}{{\textrm{ $h^{-1}$M$_{\sun}$}}}
\newcommand{\Jdir}{\ensuremath{\hat{\mathbf{J}}}}
\newcommand{\ea}{\ensuremath{\hat{\mathbf{e}}_1}}
\newcommand{\eb}{\ensuremath{\hat{\mathbf{e}}_2}}
\newcommand{\ec}{\ensuremath{\hat{\mathbf{e}}_3}}
\newcommand{\ei}{\ensuremath{\hat{\mathbf{e}}_k}}
\newcommand{\ta}{\ensuremath{\hat{\mathbf{t}}_1}}

\newcommand{\tc}{\ensuremath{\hat{\mathbf{t}}_3}}
\newcommand{\ti}{\ensuremath{\hat{\mathbf{t}}_k}}

\newcommand{\ic}{\ensuremath{\hat{\mathbf{i}}_3}}
\newcommand{\ii}{\ensuremath{\hat{\textbf{\i}}_k}}

\newcommand{\nexus}{\textsc{NEXUS+}}
\newcommand{\MMF}{\textsc{MMF}}
\newcommand{\vect}[1]{\mathbf{#1}}

\newcommand{\Req}{R_\mathrm{eq}}
\newcommand{\Rlag}{R_\mathrm{lag}}
\newcommand{\fRvir}{10 R_\mathrm{vir}}
\newcommand{\Rvir}{R_\mathrm{vir}}
\newcommand{\Rs}{R_\mathrm{s}}
\newcommand{\err}[2]{\raisebox{0.5ex}{\tiny$^{+#1}_{-#2}$}}

\definecolor{ao}{rgb}{0.0, 0.5, 0.0}

\title[Spinning masters]{Spinning masters: on the impact of tidal forces and protohalo size on early spin evolution}

 \author[L\'opez et al.]{Pablo L\'opez,$^{1,2}$\thanks{E-mail: plopez@unc.edu.ar}
 Rien van de Weygaert$^{3}$ and Manuel Merch\'an, $^{1,2}$
\\
$^{1}$Observatorio Astron\'omico de C\'ordoba, Universidad Nacional de C\'ordoba (UNC), Francisco N. Laprida 854, C\'ordoba, Argentina\\
$^{2}$Instituto de Astronom\'ia Te\'orica y Experimental, CONICET-UNC, Laprida 922, C\'ordoba, Argentina\\
$^{3}$Kapteyn Astronomical Institute, University of Groningen, PO Box 800, 9747 AD, Groningen, The Netherlands\\
}

\date{Accepted XXX. Received YYY; in original form ZZZ}

\pubyear{2022}

\begin{document}
\label{firstpage}
\pagerange{\pageref{firstpage}--\pageref{lastpage}}
\maketitle

\begin{abstract}
In this work, we explore how the size and surrounding tidal fields of dark matter protohalos at high redshift influence their angular momentum (AM) evolution. While tidal torque theory (TTT) states that AM arises from the misalignment between protohalo shape and tidal fields, it remains unclear what is the characteristic scale of the perturbations that couple with each protohalo, and its correlation with protohalo properties such as size. Moreover, although the assumptions of the TTT are assumed to hold during the linear and quasi-linear regime, cosmological simulations reveal that discrepancies between its predictions and the true AM of halos emerge earlier than expected. To address this, we analyze cosmological simulations to study tidal fields at $z=80$ using different smoothing lengths, and determine which best predicts AM under TTT. We then investigate discrepancies between predicted and actual AM across redshifts, considering the effect of evolving tidal and inertia tensors. Our results show that the early tidal field couples with the inertia tensor of protohalos on scales about half of their characteristic size and confirm that disagreements between theory and simulation \hl{emerge already at relatively early cosmic times ($z\sim10-5$)}, suggesting a systematic effect from protohalo shape interacting with the forming cosmic web.
\end{abstract}



\section{Introduction}
\label{introduction}

The acquisition of angular momentum (AM) by galaxies and dark matter (DM) halos, first studied by \citet{hoyle1949}, is crucial for galaxy formation and its link to large-scale structure (LSS). AM growth and orientation impact halo-galaxy models \citep{zentneretal2014, rodriguezetal2015, mcewenetal2018} and introduce weak lensing systematics via intrinsic alignments \citep{troxelyishak2014, hikageetal2019, fabbianetal2019, copelandetal2020}. Galaxy formation models often assume a close coupling between dark matter and baryons and conservation of angular momentum, leading to a natural link between disk size and halo size through the halo spin parameter \citep{peebles1971,fall1979,dalcantonetal1997,momaoywhite1998,somervilleetal2008,firmaniyavilareese2009,duttonetal2011}. \hl{The dependence of clustering on halo spin can be regarded as one manifestation of assembly bias \citep{shethytormen2004, gaoetal2005, wechsleretal2006, bettetal2007, gaoywhite2007}, though the origin of this correlation remains unclear \citep{monterodortaetal2024}, including possible cosmological dependencies} \citep{lazeyrasetal2021, contrerasetal2021}. AM alignments with respect to the large-scale structure may also place constraints on neutrino mass \citep{leeetal2020}.

In hierarchical structure formation, DM halos arise from collapsing primordial fluctuations, creating the gravitational framework for galaxy evolution. These structures assemble into the cosmic web, a network of nodes, filaments, walls, and voids shaped significantly by tidal fields \citep[e.g.][]{zeldovich1970, einastoetal1977, shandarinyzedovich1989, bondetal1996, hahnetal2007, weygaert&bond2008, leeetal2009, aragoncalvoetal2010, hahnetal2010, cautunetal2014, feldbruggeetal2024, kugel2024, libeskind2018}. The same tidal forces also transfer AM to halos and galaxies during collapse.

Early theoretical models proposed that AM acquisition is driven by tidal torques from the misalignment between protohalo inertia tensors and the surrounding tidal field \citep{hoyle1949, peebles1969, doroshkevich1970, efstathiouyjones1979, white1984, porcianietal2002a,porcianietal2002b}. This mechanism forms the basis of the tidal torque theory (TTT), which describes AM evolution during linear and quasi-linear structure formation. TTT provides a framework for modeling AM acquisition prior to shell crossing and helps explain how the cosmic web influences halo spin and its connection to galaxy formation (see \citealt{schafer2009} for a review).

While TTT has been successful in explaining the initial acquisition of AM and its early alignment with the LSS in a statistical sense, several critical aspects remain poorly understood. In this work, we focus on unresolved questions, such as the role of different tidal field scales in AM acquisition, the connection between these scales and protohalo properties, and the timing and nature of deviations from TTT predictions during protohalo evolution. \hl{More broadly, we ask: who are the true \emph{spinning masters}? That is, which factors are most influential in shaping the angular momentum acquired by DM halos: the spatial distribution of particles within the protohalo, the surrounding tidal field, the scale at which the latter is defined, or the time evolution of these quantities? By addressing these questions}, we aim to advance our understanding of AM evolution, its connection to present-day halo properties and, if possible, to observable galaxy properties.

TTT's effectiveness in explaining early AM distributions has made it the leading framework for this process. It successfully describes AM behavior across scales and environments, predicting how density peaks create non-Gaussian AM features while reducing variance in dense regions \citep{bardeenetal1986, heavensypeacock1988, catelanytheuns1996,schaferetal2012}. The theory reveals that AM correlations between halos stem from interactions between local inertia and large-scale tidal shear, peaking when these tensors misalign \citep{porcianietal2002a, leeypen2000}. These advances have yielded AM-tidal field parameterizations that enable direct testing through simulations and observations \citep{leeetal2020, crittendenetal2001}.

TTT studies have particularly examined spin orientation origins and LSS alignments. In idealized cases where protohalo inertia tensors and tidal fields are uncorrelated, TTT predicts spin alignment with the tidal tensor's intermediate axis \citep{catelanytheuns1996, leeypen2001}. However, simulations show strong inertia-tidal field correlations, weakening this intermediate-axis alignment and instead favoring perpendicular orientations to the tidal field's first principal axis \citep{leeypen2000, porcianietal2002b}.

A crucial finding from multiple simulation studies is that halo spin alignment depends strongly on mass. Lower-mass halos (below $\sim 10^{12} \hMsun$) typically spin parallel to their host filaments, while higher-mass halos spin perpendicular to them \citep{bailinysteinmetz2005, aragoncalvoetal2007, hahnetal2007, sousbieetal2008, codisetal2012, foreroromeroetal2014, wangykang2017, veenaetal2018, veenaetal2019, veenaetal2021, lopezetal2021, pereyraetal2020, moonylee2024}. However, observations tell a more complicated story. Different surveys using various methods have reported conflicting alignment patterns \citep[e.g.][]{navarroetal2004, leeyerdogdu2007, pazetal2008, jonesetal2010,cervantessodietal2010, tempelylibeskind2013, zhangetal2015,welkeretal2019,krolewskietal2019,bluebirdetal2020,kraljicetal2021,leeetal2023,karachentsevetal2023,rongetal2025b}, making this mass-dependent trend difficult to confirm observationally.

Numerical simulations have played a crucial role in validating and refining TTT. While the theory accurately describes AM growth in the linear regime \citep{white1984}, simulations reveal significant deviations during late time evolution \citep{sugermanetal2000, leeypen2000, porcianietal2002a, porcianietal2002b}. These discrepancies emerge as environmental effects, mergers, and vortical flows become important \citep{vitvitskaetal2002, libeskindetal2013, laigleetal2015, bettyfrenk2016, lopezetal2019, leeymoon2022}. Such results highlight the need to extend TTT with additional physics that accounts for non-linear\footnote{\hl{Hereafter we use the term “non-linear” in the loose sense common in the literature, referring to the late stages of halo evolution when the predictions of linear theory are no longer adequate to describe the dynamics of the matter. We also use it to denote processes typical of this stage, such as mergers, shell crossing, and fly-bys. In the very late phase ($z<1$), we use the term ``highly non-linear'' simply to indicate a regime where the linear approximation fails at any scale relevant for the halos in our study. In the same sense, the term ``quasi-linear'' is used to refer to intermediate redshifts ($z\sim20-4$), when we see the emergence of the first non-primordial structural patterns.
}} processes and environmental factors \citep[e.g.][]{codisetal2015, wangykang2017, neyrincketal2019, ebrahimianetal2021, moonylee2023, lopez2023}.

However, despite the extensive study of the TTT, and its effects, several critical questions remain unresolved. For instance, the early tidal field that interacts with protohalos is not scale-independent; protohalos of different sizes and shapes typically couple with tidal fields characterized by varying properties, such as overdensity and anisotropy. This raises important questions that we aim to address in this work: How do the properties of the early tidal field vary with different filter scales? More importantly, which of these scales are most relevant for the correlation between the tidal field and the inertia tensor of protohalos within the context of TTT? Furthermore, what is the connection between the relevant scales of the tidal field and the properties of protohalos, such as their size and shape? Addressing these questions is essential for refining our understanding of AM acquisition and its dependence on the initial conditions of structure formation. The key aspects of TTT that remain unclear highlight gaps in our understanding of how AM is acquired. The most challenging questions emerge from the ambiguous observational data and is related to the capacity of the model to explain the spin of galaxies, in particular their alignment with the LSS. We may identify three principal issues that involve ambiguities.

First, while TTT provides an elegant framework, simulations reveal significant deviations in DM halo AM evolution from its predictions. This suggests either theoretical incompleteness or gaps in our understanding of certain aspects, particularly the coupling scale between tidal fields and protohalo inertia tensors in Lagrangian space. While studies show that varying density field smoothing scales affect collapse times and inertia-tidal alignments \citep{leeetal2009,hahnetal2009,moonylee2024}, the conventional assumption that protohalo size determines the relevant TTT scale lacks rigorous validation. Critical unanswered questions remain: How do tidal field properties systematically vary with filtering scale? Which scales dominate tidal-inertia tensor correlations? How are these scales linked to protohalo characteristics? A thorough analysis of these scale-dependent effects is notably absent in current literature.

Additionally, the timing and nature of deviations from TTT remain unclear. While late-stage differences are often assumed based on theoretical expectations and high and low redshift comparisons, studies show deviations actually emerge progressively from early epochs \citep{lopezetal2021}. Remarkably, the primordial tidal field appears to encode information about late-time alignments \citep{cadiouetal2021,moonylee2024}. This requires comprehensive analysis of AM evolution through all growth phases to separate deviations solvable within TTT's current framework from those demanding fundamental revisions. Crucially, we must identify when protohalo AM evolution diverges from TTT predictions and how this connects to their formation history, including assembly processes, morphological evolution, and tidal field development.

Second, the TTT explains the acquisition of AM in halos, not galaxies. The mechanisms and timing involved in transferring AM from halos to galaxies remain poorly understood. This knowledge gap creates significant challenges: even with perfect understanding of halo AM acquisition, baryonic processes during galaxy formation may overwrite primordial tidal imprints, while non-linear evolution also alters halo orientations. The combined effect means observed galaxy spins could reflect neither their initial configurations nor their host halos' current states. Hence, baryons may potentially inherit either the halo's early TTT-predicted AM, a modified version shaped by intermediate processes, or properties completely divorced from the original tidal torque scenario. Resolving this baryon-halo decoupling timeline is essential for connecting TTT to observable galaxies.

Lastly, observational studies present a complex and often contradictory picture of galaxy spin alignments. Clear trends emerge in some surveys: low-mass spirals typically align parallel to filaments, while high-mass systems and S0 galaxies show perpendicular orientations \citep{kraljicetal2021,welkeretal2019,tempelylibeskind2013}, with similar mass-dependent flips observed around void centers \citep{leeetal2023}. These patterns further depend on morphology, with red early-types exhibiting alignments absent in their blue/green counterparts \citep{rongetal2025b}. Yet other studies find no significant alignments, whether in Local Volume spirals \citep{karachentsevetal2023} or MaNGA galaxies \citep[][aside from tentative H$\alpha$ signals]{krolewskietal2019}. The strength of measured correlations varies dramatically, from strong tidal alignments \citep{leeyerdogdu2007,lee2011} to weak or undetectable signals in galaxy pair studies \citep{zhangetal2015,cervantessodietal2010}, leaving the observational status of spin-LSS connections fundamentally unresolved.

Attempts to reconcile some of these results within the TTT framework have led some researchers to warn against the theory turning from an empirical into a `vampirical'' hypothesis, meaning it resists falsification because its predictions can always be reinterpreted to fit the data \citep{andraeyjahnke2011}. Although this problem lies beyond the scope of our work, we hope our findings help bridge the gap between TTT predictions at high redshift and observations in the local Universe. By shedding light on the nature of deviations from the theory, our results may also provide insights into whether inconsistencies can be reconciled through reinterpretation or require a deeper revision of the theoretical framework.

Hence, this paper is organized as follows: Section \ref{sec:TTT} reviews TTT's foundational concepts and standard implementation. Section \ref{sec:methods} details our simulation approach for analyzing the early tidal field at multiple filter scales, protohalo properties, and cosmic web geometry. We then present our results in three stages: Section \ref{sec:early_tf} characterizes the scale-dependent properties of the early tidal field; Section \ref{sec:TTT_predictions} examines how different tidal scales couple with protohalo inertia tensors to generate AM; and Section \ref{sec:spin_evolution} tracks AM evolution from linear theory to present-day halos, including: (1) particle-specific AM distributions, and (2) temporal variations in tidal fields and inertia tensors. Finally, in Section \ref{sec:summary}, we summarize our findings and the main conclusions of this study.

\section{Brief outline of the TTT}
\label{sec:TTT}

The TTT describes the growth of AM at early times, within the context of a Friedmann-Lema\^{i}tre-Robertson-Walker (FLRW) universe which expands isotropically over cosmic time $t$ according to the scale factor $a(t)$, with an associated redshift $z(t)=a^{-1}(t)-1$. The model adopts a Lagrangian framework, treating the individual mass elements as distinct particles and defining their ``initial positions'' at a sufficiently early time $t_0$, or \textit{Lagrangian coordinates}, as $\mathbf{q}=\mathbf{x}(t_0)$, where $\mathbf{x}(t)$ are the usual comoving coordinates, related to the physical coordinates $\mathbf{r}(t)=a(t)\mathbf{x}(t)$. In this context, the TTT assumes that, prior to shell crossing, the evolution of the particles can be well described using the Zel'dovich approximation \citep{zeldovich1970}. It also assumes that, given a region $V_\mathrm{L}$ that will end up forming a halo at present time, i.e. a \textit{protohalo}, the gravitational potential around its center of mass can be approximated by its second-order Taylor expansion when the associated density field is smoothed over an appropriate scale. 

Hence, consider a protohalo at $t_0$ that is characterized by its inertia tensor: 
\begin{equation}
    \mathcal{I}_{ij} = \bar{\rho}_0 a^3_0 \int_{V_L} q_i q_j d^3q,
\end{equation}
where $\bar{\rho}_0$ and $a_0$ are, respectively the average background matter density and the value of the scale factor at this moment. Now consider that this protohalo is subjected to a tidal field associated to a deformation tensor:
\begin{equation}
    \mathcal{D}_{ij} = \frac{\partial^2\phi_R}{\partial q_i \partial q_j},
\end{equation}
where $\phi_R$ is the gravitational potential appropriately smoothed on a scale $R$. This is of course related to the smoothed density field $\rho_R$ via the Poisson equation:
\begin{equation}
    \nabla^2\phi_R=4\pi G \bar{\rho}_0\delta_R,
\end{equation}
where $G$ is the gravitational constant and $\delta_R=\rho_R/\bar{\rho}_0-1$ is the smoothed overdensity. Under these conditions, according to the TTT, the early evolution of the $i$-th component of the protohalo's AM is given by \citep{white1984}:
\begin{equation}
    \label{eq:TTT}
    J_i(t) = a^2(t) \dot{D}(t) \epsilon_{ijk} \mathcal{D}_{jl} \mathcal{I}_{lk},
\end{equation}
where $D(t)$ is the linear growth factor describing the evolution of density perturbations, the dot denotes derivation with respect to the cosmic time and $\epsilon_{ijk}$ is the fully antisymmetric Levi-Civita tensor (for a more comprehensive explanation of TTT, see, for example, Section 2 of \citealt{lopezetal2019} and references therein).

\begin{figure*}[H]
    \includegraphics[width=2\columnwidth]{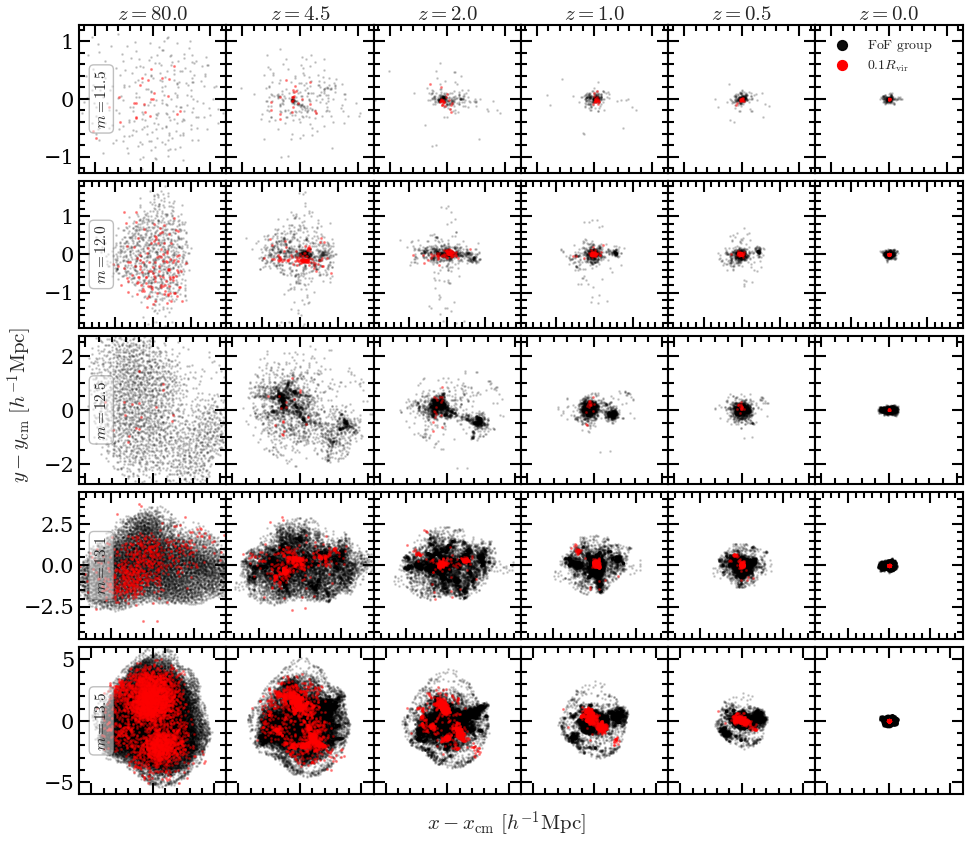}
    \caption{Five examples of the evolution of protohalos, with increasing mass from top to bottom. Each column, from left to right, shows the particle distribution at $z=80, 4.5, 1.5, 0.5$ and $0$, respectively. The black dots represent the full FoF group, while the red dots show the particles that reside within $0.1\Rvir$ at present time (see Section \ref{sec:spin_evolution}). Notice that the axes that correspond to each protohalo are scaled differently.}
    \label{fig:protohalos}
\end{figure*}

This equation indicates that, during the early stages of structure formation, the AM arises from the misalignment between the shape of the protohalo and the principal directions of expansion or collapse associated to its surrounding tidal field \emph{appropriately smoothed}. The ambiguity of this expression is deliberate: in general, it is not prescribed what is the scale of the tidal field that couples with the inertia tensor to produce AM. It is usually assumed that corresponds roughly to the size of each protohalo. However, \hl{(1) in Lagrangian space, protohalo boundaries are highly irregular and \hl{may be} non–simply-connected (see Figure \ref{fig:protohalos}), so their size and shape can only be established a posteriori from the particle membership assigned by a halo finder; moreover, while such criteria are well defined in Eulerian space at the time of identification, their mapping to the Lagrangian region is less direct and introduces additional arbitrariness}, and (2) the amplitude and orientation of the tidal field can change with the smoothing scale both below and above the characteristic size of each protohalo. Hence, even with an unequivocal definition of the protohalo size $R_\mathrm{h}$, the resulting tidal field, and thus the TTT predictions, can be highly sensitive to the smoothing scale, differing substantially when the density field is smoothed over, e.g., $R_\mathrm{h}/2$, $R_\mathrm{h}$, or $2R_\mathrm{h}$.

The uncertainty in this aspect is surprising when we consider that the eigenvalues of the corresponding tidal tensor, which change with the smoothing scale, determine the overdensity and anisotropy of the associated region and are strongly connected to protohalo properties in Lagrangian space \citep{Ludlow2014,borzyszkowskietal2014}. It is therefore important to more clearly specify the relationship between the size of a protohalo and the characteristic scale of the tidal field that couples with its shape to produce AM.

Another important feature of Eq. \eqref{eq:TTT} is that the time dependence of $J_i(t)$ is determined exclusively by the \textit{isotropic} factor $a^2(t) \dot{D}(t)$. This implies that all components of the AM grow at the same rate. Consequently, while the magnitude of the AM changes over time, its direction remains constant, provided the assumptions of the model hold. This aspect of the TTT has significant implications: while it predicts preferred directions for the AM that depend on the correlation between $\mathcal{I}_{ij}$ and $\mathcal{D}_{ij}$, it does not account for changes in spin orientation. Many studies that implement this framework to estimate AM directions relative to the LSS \citep[e.g.,][]{leeypen2000, porcianietal2002b}, or to explain secondary trends such as the mass dependence of spin-filament alignment \citep{codisetal2015}, often neglect the time-dependent factor and focus primarily on the properties of both the primordial inertia and tidal tensors and their correlation. The variation of these quantities with time can be accounted for when considering higher-order Lagrangian perturbation theory \citep{catelanytheuns19962LPT} but, to our knowledge, this has not been applied to the early evolution of the AM orientation.


\section{Methods}
\label{sec:methods}

\subsection{The simulation, halos and protohalos}
\label{sec:methods:simulation}

In this work we use the code \textsc{gadget 2} \citep{springelgadget2005} to simulate the cosmological evolution of $1600^3$ dark matter particles in a periodic cube of side $400\hMpc$, with a mass per particle $m_{\mathrm{p}}= 1.18219 \times 10^{9}\hMsun$. This simulation consists of 205 outputs between $z=80$, when the initial conditions were generated, and $z=0$, which corresponds to the present time. Cosmological parameters were taken from the Planck Collaboration results \citep{plankcollaboration2018}: matter density $\Omega_{\mathrm{m}}=1-\Omega_{\Lambda}=0.315$, Hubble constant $H_{0}=67.4 \kms \Mpc^{-1}$ and normalization parameter $\sigma_{8}=0.811$. Dark matter halo identification was performed at $z=0$ using a standard \textit{Friends of Friends} (FoF) algorithm with percolation length $l=0.17\bar{\nu}^{-1/3}$, where $\bar{\nu}$ is the mean particle number density. In order to avoid biases in the determination of dynamical properties due to the low number of particles \citep[see e.g.][]{pazetal2006,bettetal2007}, our analysis was carried out discarding FoF groups with less than $250$ particles. The result is a sample of about $6.5\times10^5$ halos with masses {$M_\mathrm{FoF}\geqslant 3\times 10^{11}\hMsun$}.

In order to study the effect of the early tidal field over the origin and evolution of the AM, we adopt a Lagrangian approach. Once we identify the FoF groups at $z=0$, we follow their particles back in time to the initial conditions of the simulation. At each snapshot, we compute the AM of the corresponding protohalo:
\begin{equation}
      \vect{J}(z) = m_\mathrm{dm} \sum_{p = 1}^{N_\mathrm{h}}
      \vect{r}_{p}(z) \times \vect{v}_{p}(z),
\end{equation}
where $N_\mathrm{h}$ is the number of particles in the group and $\vect{r}_{p}(z)$ and $\vect{v}_{p}(z)$ represent the position and velocity vectors of the $p$-th particle with respect to the center of mass of the protohalo at redshift $z$. We also calculate its inertia tensor:
\begin{equation}
      I_{ij}(z) = \frac{1}{N_\mathrm{h}} \sum_{p = 1}^{N_\mathrm{h}}
      r_{pi} r_{pj}(z),
\end{equation}
and the corresponding eigenvalues $i_k$ and eigenvectors $\ii$, with $k=1,2,3$.

In addition to analyzing the full FoF groups, we track the particles that, by $z=0$, reside within fixed fractions of the virial radius ($0.1\Rvir$ and $0.5\Rvir$) of each halo (see Section \ref{sec:spin_evolution}). By comparing these \emph{inner particles} to their outer counterparts, we aim to understand whether the AM assembly is driven differentially (e.g., if inner particles experience distinct tidal torques or non-linear effects) or if the entire Lagrangian region evolves coherently. This distinction is important for understanding how the AM of the final halo relates to its initial conditions.

Figure \ref{fig:protohalos} presents the Lagrangian evolution of five representative FoF groups (black dots) and their inner $0.1\Rvir$ particles (red dots) at redshifts $z=80,4.5,1.5,0.5$, and $0$. The full FoF groups exhibit diverse initial configurations, ranging from contiguous to fragmented spatial distributions, particularly for higher-mass halos. This morphological variety demonstrates the complexity of characterizing Lagrangian regions. The inner particles show similarly irregular distributions, either scattered throughout the entire volume or partially clustered in subregions.

The evolutionary sequence reveals distinct phases of collapse. By $z=4.5$, particles remain broadly distributed, with only faint filamentary structures emerging (a pattern reminiscent of large-scale structure formation\hl{; see \citealt{neyrincketal2025} for a detailed analysis of such ``halo root systems''}). The distribution becomes increasingly clumpy by $z=1.5$, with most inner particles forming an offset core that becomes prominent by $z=0.5$.

While these inner particles must by definition collapse into the halo core by $z=0$, their early evolutionary paths differ substantially. Some maintain coherent substructures throughout the collapse, while others become thoroughly mixed with external material. This contrast emphasizes the importance of separately analyzing the angular momentum contributions from inner and outer regions to fully understand halo formation.

Finally, we need to characterize the large-scale structure at present time. To this end, \hl{we use the code \nexus{}} \citep{cautunetal2013} at $z=0$, which establishes the type of region in which each halo is embedded through an analysis of the Hessian of the smoothed local density field at different scales. In this way, it identifies not only the cosmological environment (void, wall, filament or node), but also the preferred directions of collapse and expansion of matter, $\hat{\vect{e}}_k$, with $k=1,2,3$. In the case of filamentary regions, for example, $\hat{\vect{e}}_3$ can be associated with the direction of the main axis or \emph{spine}. A more detailed description of \nexus{} can be found on Appendix \ref{app:nexus}.

\subsubsection{The size of protohalos}
\label{sec:methods:protohalo_size}

As shown in Figure \ref{fig:protohalos}, protohalos are usually associated to irregular regions in Lagrangian space. The size and shape of these regions are not unequivocally defined, and hence we have to make a decision on how to characterize them. In this work, we explore three different methods for measuring the size of protohalos at $z=80$, each associated with a distinct aspect or evolutionary stage of the particle distribution.

A straightforward approximation to the size of a protohalos in Lagrangian space is an homogeneous sphere of density $\bar{\rho}_0$ that would contain the mass $M$:
\begin{equation*}
R(M)=\left[M\bigg/\left(\frac{4}{3}\pi\bar{\rho}_0\right)\right]^{1/3}.
\end{equation*}
For our purposes, we define the \emph{Lagrangian radius}, $\Rlag=R(M_\mathrm{FoF})$. This quantity depends solely on the mass of the protohalos, which is, by definition, constant over time. When considered as a smoothing scale, $\Rlag$ has been found to maximize the correlation between the early tidal field around protohalos and their inertia tensor \citep{leeetal2009}, while a related quantity, $R(M_\mathrm{FoF}/2)$ has been shown to maximize the correlation between the peak height associated with the smoothed density field and the formation time of the corresponding halos \citep{hahnetal2009}. 

Our second measure of protohalo size is the \emph{equivalent radius}, \hl{which we define as the radius of a sphere that is equal in volume to the ellipsoid associated with the eigenvalues of the inertia tensor. Hence, given the semiaxes $a,b,c=\sqrt{\mathrm{i}_k}$ for $k=1,2,3$, the equality $\frac{4}{3}\pi \Req^3 = \frac{4}{3}\pi\,abc$ allows to compute the equivalent radius as:}
\begin{equation*}
\Req =(abc)^{1/3}.
\end{equation*}
This quantity is computed using the particle distribution of each protohalo at $z=80$, i.e., at the same time when we analyze its surrounding tidal field. 

\begin{figure}
    \includegraphics[width=\columnwidth]{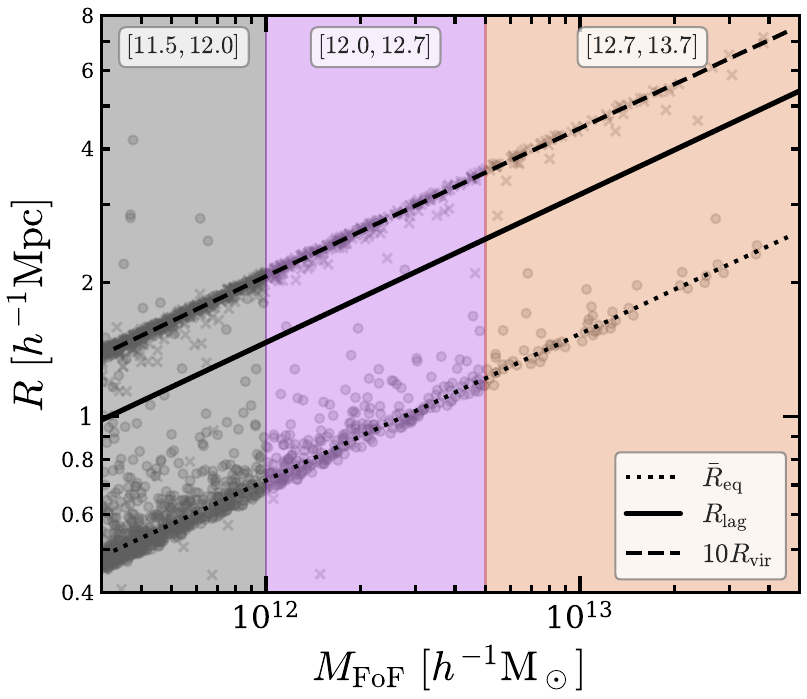}
    \caption{Protohalo size as a function of their mass. The dot, solid and dashed lines show, respectively, the median of the equivalent radius ($\Req$), the Lagrangian radius ($\Rlag$), and ten times the virial radius ($\fRvir$) for protohalos of different mass. While the value of $\Rlag$ is computed exactly from the halo mass, $\Req$ and $\fRvir$ present some scatter which can be seen as the dispersion of the gray circles and crosses. The colored regions indicate the three mass bins we use in our analysis.}
    \label{fig:Rlag_vs_M}
\end{figure}

\hl{Finally, we have $\fRvir$, i.e. ten times the virial radius of each present-day halo. While $R_\mathrm{vir}$ is measured at $z=0$, $\fRvir$ is treated as a comoving size that serves as a fixed scale at any redshift. The factor multiplying $R_\mathrm{vir}$ aims to take into account that, at early times, the protohalo particle distribution is significantly more extended in comoving coordinates than the corresponding virialized object at $z=0$. The value of this factor has been chosen such that $\fRvir$ represents a scale typically larger than $\Rlag$ at $z=80$.} \citet{moonylee2024} found that calculating the tidal field using a similar filter length, $8 R_\mathrm{vir}$, yields a universal correlation with the inertia tensor, independent of the mass of each protohalo. Moreover, they found that $8 R_\mathrm{vir}$ establishes an upper limit beyond which the so-called ``type II spin transition'' is not observed, i.e., the transition of $\Jdir$ from $\perp\ta$ to $\parallel\ta$, where $\ta$ represents the first axis of the tidal field. Thus, the value $\fRvir$ allows us to test these results and compare our conclusions.

Figure \ref{fig:Rlag_vs_M} shows the relation between halo mass and the three different measures of the protohalo size at $z=80$: the equivalent radius ($\Req$), the Lagrangian radius ($\Rlag$) and ten times the virial radius ($\fRvir$). Notice that, whereas the value of $\Rlag$ is computed exactly from the halo mass, $\Req$ and $\fRvir$ present some scatter which can be seen as the dispersion of the gray circles and crosses, respectively. Naturally, the protohalo size tends to increase with halo mass. The equivalent radius spans a range of values such that its median goes from $\sim0.5$ to $2.6\hMpc$ in the mass range $[3\times10^{11}-5\times10^{13}]\hMsun$. Typically, for each protohalo, $\Req$ is about half of the value of $\Rlag$ and $\sim 1/3$ of $\fRvir$. 

In order to consider the possible dependence of our results with halo mass, we split our halo population in three subsamples, shown in Figure \ref{fig:Rlag_vs_M} as coloured regions. With\footnote{\hl{Hereafter, we use the convention $\log\equiv\log_{10}$.}} $m=\log\left[M_\mathrm{FoF}/(\hMsun)\right]$, we define a low mass subsample within the range $m=[11.5,12.0]$, a subsample of intermediate mass $m=[12.0,12.7]$ and, finally, a high mass subsample with masses in the range $m=[12.7,13.7]$. These ranges have been chosen such that a well known trend in spin alignment at $z=0$ can be easily identified: we expect the spin of the low (high) mass subsample to be preferentially aligned (perpendicular) to the present day filaments.

 \subsection{The tidal field at different scales}
\label{sec:methods:tidal_field}

One of the key ingredients of the TTT is the tidal field associated with the position of each protohalo during the linear regime. However, it is commonly stated that, in order to eliminate the contribution of the protohalo itself to the perturbations responsible for its torquing, the potential field must be properly smoothed at protohalo scales before computing its second derivatives. This ensures not only that the resulting torque is \emph{external} but also that the velocity flow is laminar. Although different methods have been adopted to perform this analysis, it is not clear which scale is most appropriate to correctly implement the TTT. In general, it is assumed that this scale is of the order of $\Rlag$, but applying a Gaussian or top-hat filter of this characteristic size in Fourier space \hl{suppresses not only the contribution from substructures and fluctuations within each protohalo}, but also that of all perturbations associated with similar or smaller wavelengths.

In order to analyze the dependence of the TTT predictions on the characteristic scale of the tidal field, we carry out a systematic study of the effect that different smoothing lengths, $\Rs$, have on the tidal tensor $\mathbf{T}(\Rs)=T_{ij}(\Rs)$, through the following method. First, we determine the overdensity field $\delta(\vect{x})$ on a cubic grid of $1024^3$ cells, by means of a \textit{cloud in cell} (CIC) interpolation. We then use \nexus{} to compute, over this grid, the tidal tensor $T_{ij}(\mathbf{x})=\partial^2 \Phi(\mathbf{x})/\partial x_i\partial x_j$, where $\Phi(\mathbf{x})=\phi(\mathbf{x})/(4\pi G \bar{\rho})$ and $\bar{\rho}$ is the mean comoving background density, by inverting in Fourier space the Poisson equation $\nabla^2 \Phi=\delta(\mathbf{x})$. In this process we apply a series of Gaussian filters on the density field of the form $G(R_\mathrm{s})=e^{-k^2R_\mathrm{s}^2/2}$, where $R_\mathrm{s}$ takes $30$ logarithmically equispaced values between $R_\mathrm{s}=0.5\hMpc$ and $30\hMpc$. This allows to produce multiple smoothed versions of the tidal field, $T_{ij}(\vect{x},R_\mathrm{s})$. We then interpolate these fields in coordinate space to the location $\vect{x}_\mathrm{h}$ of each halo. In this way we obtain a halo-centric catalogue of tidal tensor estimates with different smoothings, $T_{ij}(\vect{x}_\mathrm{h},R_\mathrm{s})$. Finally, in order to determine the preferred expansion or collapse directions, we diagonalize each tidal tensor and compute the corresponding eigenvalues and eigenvectors, $\lambda_k(\vect{x}_\mathrm{h},R_\mathrm{s})$ and $\hat{\vect{t}}_k(\vect{x}_\mathrm{h},R_\mathrm{s})$, with $k=1,2,3$, respectively. In what follows we will always assume that the eigenvalues and eigenvectors correspond to the halocentric tidal field, and therefore we will omit the dependence on the position of each proto-halo, $\vect{x}_\mathrm{h}$.

\begin{figure*}
    \includegraphics[width=2\columnwidth]{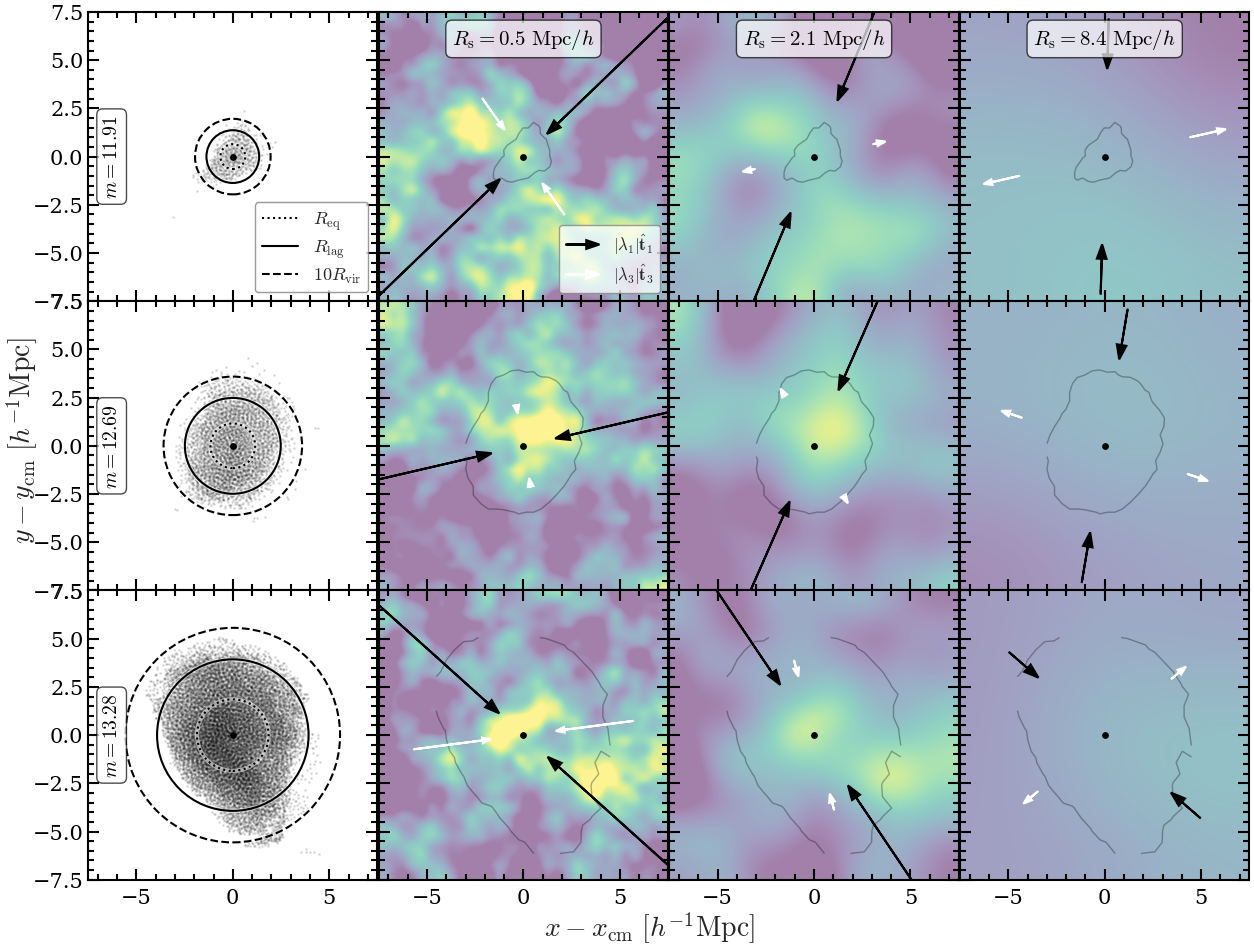}
    \caption{Three examples of protohalos at $z=80$, with increasing mass from top to bottom. The first column shows the particle distribution and the three measurements of protohalo size: the equivalent radius ($\Req$), the Lagrangian radius ($\Rlag$), and ten times the virial radius ($\fRvir$). The other three columns show the density field around each protohalo after aplying a gaussian filter with smoothing lengths $\Rs/(\hMpc)=0.5$, $2.1$ and $8.4$. The contours reproduce the protohalos while the arrows show the projected first and third axis of the resulting tidal tensor.}
    \label{fig:halos_ejemplo}
\end{figure*}

Figure \ref{fig:halos_ejemplo} illustrates the result of this procedure. Each row corresponds to the density field in the vicinity of a protohalo at redshift $z=80$. From top to bottom, the protohalos have masses $M_\mathrm{FoF}=8.1\times10^{11}$, $5.7\times10^{12}$ and $1.9\times10^{13}\hMsun$, respectively. In each case, the first column shows the particle distribution of the protohalo along with three different measurements of its size, $\Req$, $\Rlag$ and $\fRvir$, with dotted, continuous and dashed circles, respectively. The other three columns columns show the surrounding density field smoothed with gaussian kernels with radii $Rs/(\hMpc)=0.5,2.1$ and $8.4$, from left to right. The direction of the arrows indicates the orientation of the first and third axes of the smoothed tidal field, $\hat{\vect{t}}_1(R_\mathrm{s})$ and $\hat{\vect{t}}_3(R_\mathrm{s})$, respectively. The sign of the corresponding eigenvalues, $\lambda_1(R_\mathrm{s})$ and $\lambda_3(R_\mathrm{s})$, determines whether the matter is collapsing or expanding in that direction, whereas their absolute value is proportional to the length of the arrows.

In general, we observe that the magnitude of the eigenvalues approaches zero as the smoothing scale increases. For small values of $\Rs$, all of the eigenvalues are typically positive, indicating that matter is mostly collapsing at scales below the size of the protohalos. However, one or more eigenvalues can become negative for larger values of $\Rs$ if the protohalo is located in an underdense region or within a significantly anisotropic environment. The direction of the principal axes of the surrounding tidal tensor also changes as $\Rs$ varies. This highlights that choosing an appropriate scale for implementing Eq. \eqref{eq:TTT} is not trivial, since the correlation with the inertia tensor will also depend on $\Rs$.

\section{The early tidal field}
\label{sec:early_tf}

\subsection{Properties of the tidal field at $z=80$}
\label{sec:early_tf:properties}

\begin{figure*}
    \includegraphics[width=1.8\columnwidth]{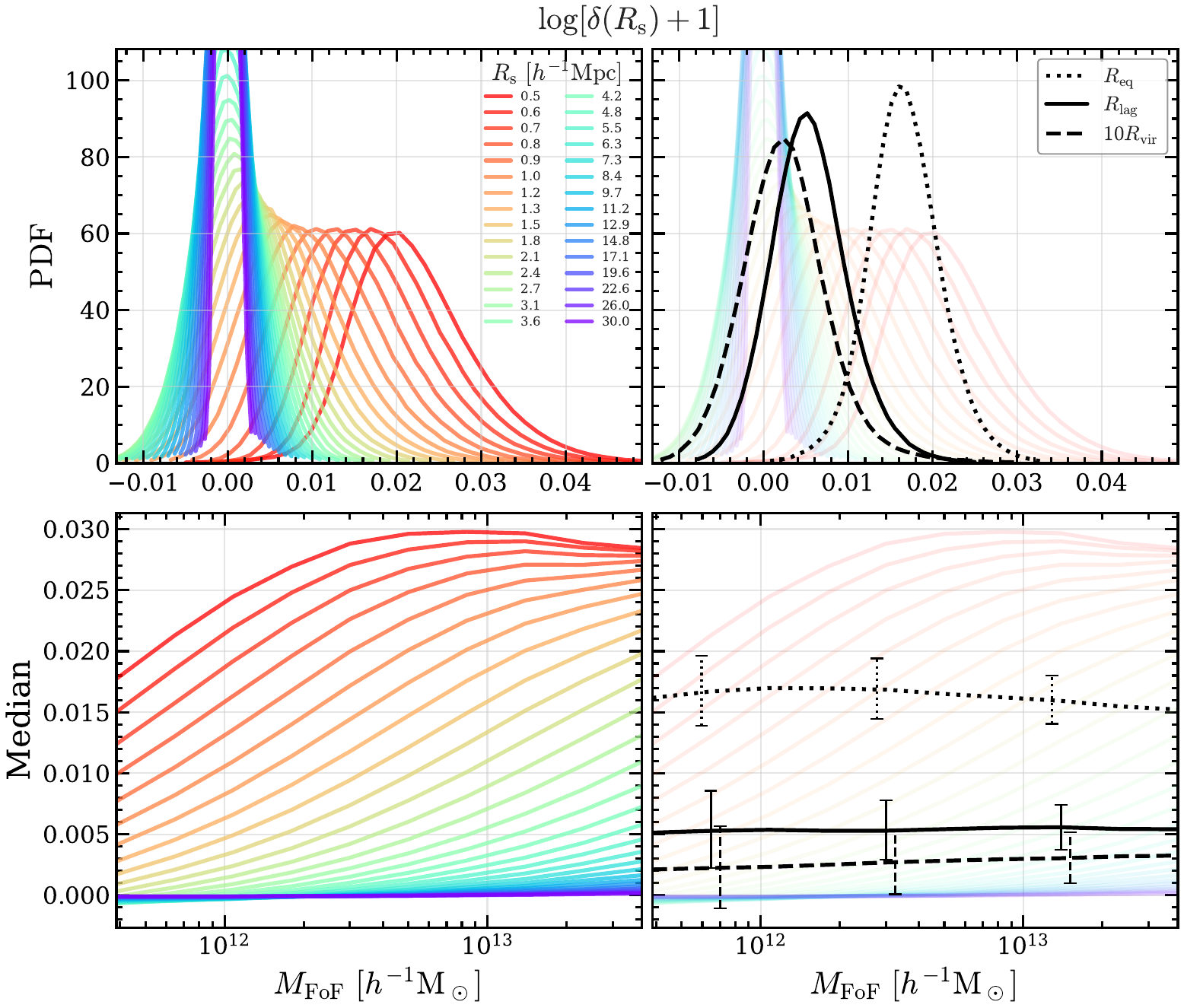}
    \caption{Distribution and median of the density contrast, $\delta$, associated with the smoothed tidal fields. The top row shows the PDFs for fixed smoothing scales $\Rs$ (left) and for protohalo size-dependent filters corresponding to $\Req$, $\Rlag$ and $\fRvir$ (right). The bottom row shows the median values of $\log(\delta+1)$ as a function of protohalo mass, $M_\mathrm{FoF}$, for the same filter scales. Error bars in the bottom-right panel indicate the interquartile range for three different mass bins.}
    \label{fig:delta}
\end{figure*}

In this work we analyze the initial tidal field in terms of the eigenvalues of the corresponding tidal tensor. Two properties are useful for characterizing it: the density contrast and the level of anisotropy. The formation and evolution of structure depends on how these quantities couple with the density field. Given a tidal tensor $\mathbf{T}(\Rs)$ computed using a gaussian filter of scale $\Rs$, the associated overdensity is simply: 
\begin{equation}
\delta = \lambda_1+\lambda_2+\lambda_3,
\label{eq:delta}
\end{equation}
where $\lambda_k$ are the eigenvalues of $\mathbf{T}(\Rs)$. The value of $\delta$ will depend, thus, in the smoothing scale $\Rs$. 

The eigenvalues of the tidal tensor also allow to measure the level of anisotropy in the gravitational pull to which matter is subject, which is expected to play an important role in the spin direction predicted by TTT. A natural approach in order to characterize anisotropy is to examine the rotational invariants of the tidal tensor $\mathbf{T}(\Rs)$ beyond its trace $\delta$. A useful quantity in this context is $q^2$ \citep{heavensypeacock1988,catelanytheuns1996}, often referred to as \emph{tidal shear}. This measure is defined as:
\begin{equation}
\begin{split}    
q^2 & = T_1^2-3T_2 \\
     & = \frac{1}{2} \left[ (\lambda_2-\lambda_3)^2 + (\lambda_3-\lambda_1)^2 + (\lambda_1-\lambda_2)^2 \right],
\end{split}  
\label{eq:q2}
\end{equation}
where $T_1=\delta$ and $T_2=\lambda_2\lambda_3+\lambda_3\lambda_1+\lambda_1\lambda_2$ are the first two rotational invariants of $\mathbf{T}(\Rs)$. As shown by \citet{shethytormen2002}, an interesting aspect of this parameter is that its distribution is expected to be independent of $\delta$. 

In Figure \ref{fig:delta} we present the distribution of $\log(\delta+1)$ for all protohalos (upper row) and the variation of the median of each distribution as a function of protohalo mass (bottom row). The left-hand panels correspond to the results when computing the tidal tensor using $30$ equally log-spaced smoothing scales $\Rs$ between $0.5\hMpc$ and $30\hMpc$. Each value of $\Rs$ is represented by a coloured curve. With no surprise, the PDFs show that the expected value of $\delta$ typically approaches zero as $\Rs$ increases, while the dispersion narrows down. In other words, the overdensity associated with the tidal field becomes smaller when we consider longer wavelength modes. This is true for the general trend of the colored curves, but the specific values of $\delta$ for a given smoothing scale varies significantly around protohalos of different mass, especially when we consider small values of $\Rs$. This can be seen in the bottom-left panel: for $\Rs\sim1\hMpc$, for example, the overdensity varies from $\log(\delta+1)\sim0.006$ in the low mass extreme to $\sim0.026$ for high mass protohalos. In general, a fixed smoothing scale yields higher values of $\delta$ for more massive protohalos, except for $\Rs<0.8\hMpc$ and $M_\mathrm{FoF}>10^{13}\hMsun$. Conversely, for any fixed mass, the tendency is that the overdensity increases as $\Rs$ decreases.

  \begin{figure*}
	\includegraphics[width=1.8\columnwidth]{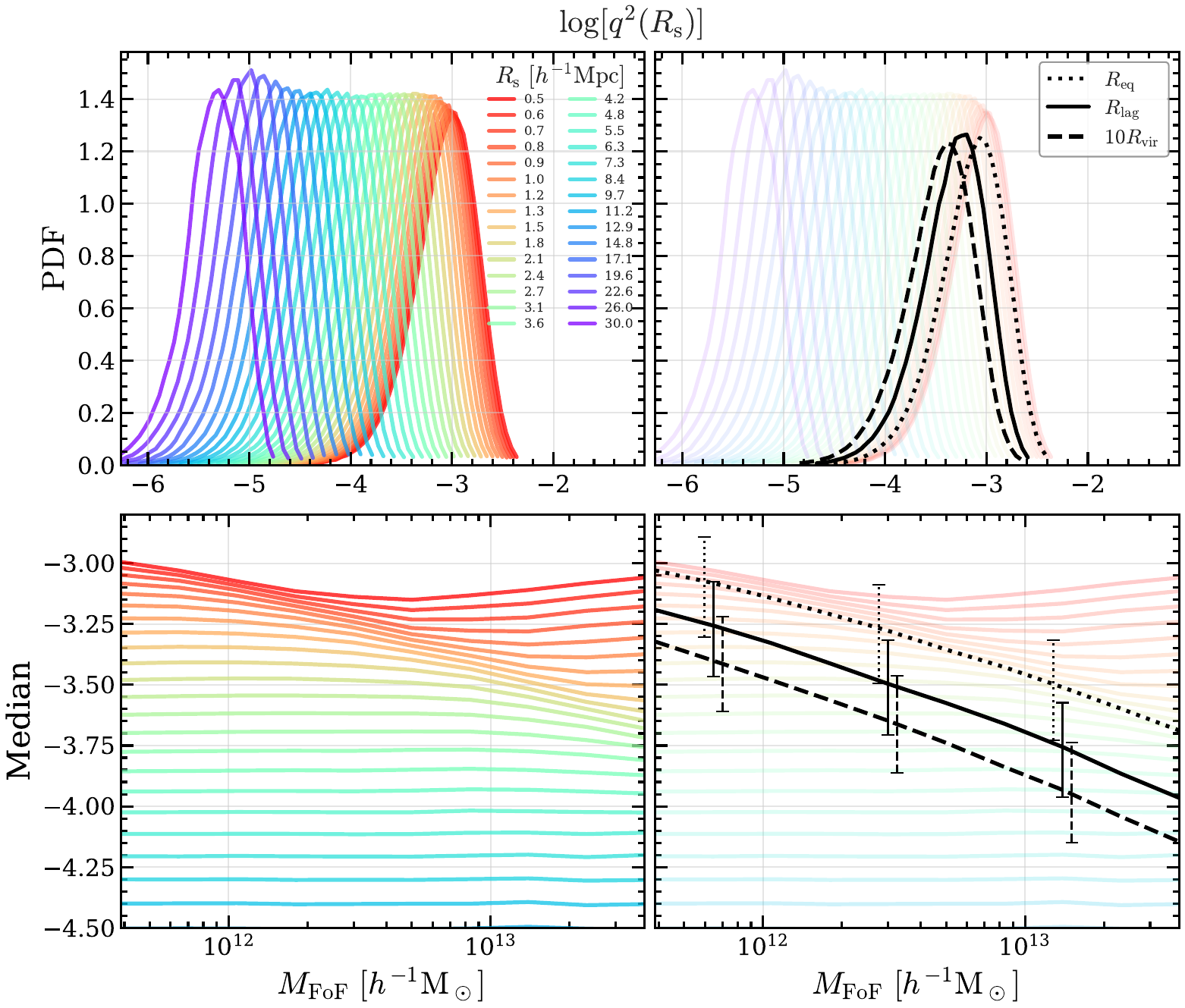}
   \caption{Distribution and median of the tidal shear, $\log(q^2)$, which quantifies the anisotropy of the smoothed tidal fields. The top row shows the PDFs for fixed smoothing scales $\Rs$ (left) and for protohalo size-dependent filters corresponding to $\Req$, $\Rlag$ and $\fRvir$ (right). The bottom row shows the median values of $\log(q^2)$ as a function of protohalo mass, $M_\mathrm{FoF}$, for the same filter scales. Error bars in the bottom-right panel indicate the interquartile range for three different mass bins. \hl{In the bottom panels, curves corresponding to the highest $\Rs$ are omitted, as they show nearly constant values with negligible mass dependence and would obscure the behavior observed in the low-$\Rs$ curves.}}
   \label{fig:qr2}
 \end{figure*}

So far, the behaviour of the tidal field smoothed in different scales is as expected. What happens when, for each protohalo, we consider a value of $\Rs$ that takes into account its size? This is shown in the right-hand panels of Figure \ref{fig:delta}, where we present with black dotted, continuous and dashed lines the resulting overdensities when we compute the tidal tensor setting, for each protohalo, $\Rs=\Req,\Rlag$ and $\fRvir$, respectively. When looking at the PDFs, we observe that the equivalent radius typically yields the higher overdensities, which is a consequence of $\Req$ being the smaller measure of protohalo size. However, the most interesting feature of this analysis can be seen in the relation between $\log(\delta+1)$ and the protohalo mass: the median of every distribution becomes practically mass-independent. In effect, the black curves in the bottom-right panel present little or no variation at all when we compare their value across the entire mass range of protohalos. As already noted, $\Req$ produces the stronger overdensities ($0.017\err{0.003}{0.003}$, $0.017\err{0.003}{0.003}$ and $0.016\err{0.002}{0.002}$, from left to right in the position of the error bars), whereas $\Rlag$ ($0.005\err{0.003}{0.003}$, $0.005\err{0.002}{0.002}$ and $0.006\err{0.002}{0.002}$) and $\fRvir$ ($0.003\err{0.003}{0.003}$, $0.003\err{0.003}{0.003}$ and $0.002\err{0.002}{0.002}$) yield more homogeneous density distributions.

For the equivalent radius, $\Req$, this result is less surprising, since it is computed using the particle distribution at $z=80$ and thus already contains information about the underlying density field. However, $\Rlag$ and $\fRvir$ are obtained from $z=0$ properties such as the total halo mass and the virial radius, with no prior knowledge about the particular spatial distribution of the protohalo particles. Hence, these results show that, despite the diversity of masses, shapes and evolutionary tracks of protohalos, it is possible to find characteristic overdensities around protohalos based on their present-day properties.

  \begin{figure*}
	\includegraphics[width=2\columnwidth]{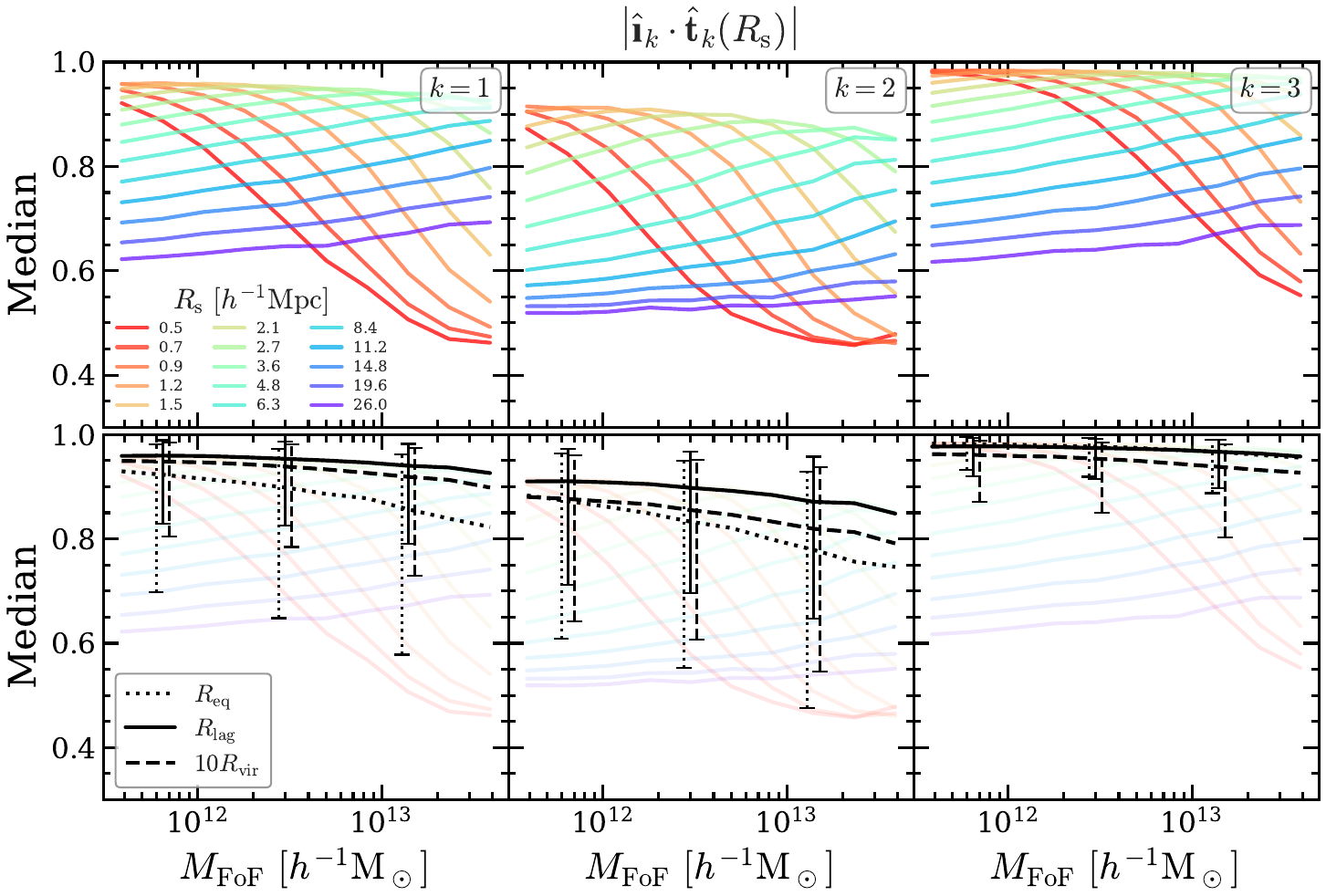}
   \caption{Median alignment between the principal axes of the inertia tensor of protohalos and the smoothed tidal tensor, $|\ii \cdot \ti(\Rs)|$, for $k=1,2,3$, as a function of protohalo mass, $M_\mathrm{FoF}$. The top row shows the results for fixed smoothing scales $\Rs$, while the bottom row presents the alignment for size-dependent filters corresponding to $\Req$, $\Rlag$ and $\fRvir$. Error bars in the bottom row indicate the interquartile range for three different mass bins.}
   \label{fig:medalign_TFvsI}
 \end{figure*}
 
In Figure \ref{fig:qr2}, we present the distribution of \hl{$q^2$, i.e. the tidal shear}. As in the previous figure, the upper panels show the PDF of $\log(q^2)$ for all protohalos and different smoothing scales, whereas the bottom panels display the variation of the median of each distribution as a function of protohalo mass. Focusing on the top-left panel, we observe that the tidal field becomes, in general, more isotropic as larger values of $\Rs$ are chosen, i.e., there is a systematic shift toward lower values of $q^2$. This behavior results from the Gaussian filter suppressing the shorter-wavelength perturbations of the primordial density field. When we take into account the protohalo mass (bottom-left panel), the simplest trend is observed for large values of $\Rs$. At any fixed smoothing length beyond $\sim 4\hMpc$, the median of $\log(q^2)$ becomes independent of the protohalo mass, showing that this scale represents somewhat of an upper bound for the correlation between the tidal anisotropy and the local overdensities.
For values of $\Rs$ below $\sim4\hMpc$, however, the slope of each coloured curve changes significantly with $M_\mathrm{FoF}$. In particular, for each smoothing scale there is a range of values of $M_\mathrm{FoF}$ within which the median of $\log(q^2)$ descends abruptly. For $\Rs<1\hMpc$ the curves reach a minimum and then start increasing. The mass where this change of slope occurs is higher for larger values of $\Rs$. 

When considering protohalo size-dependent smoothing lengths (right-hand panels), we observe that the anisotropy typically increases from the inner to the outer measurements of protohalo size. In effect, the upper-right panel shows that, when filtered successively with $\Rs=\Req,\Rlag$ and $\Rvir$, the PDF of the tidal shear shift towards lower values. This can also be seen in the bottom-right panel: at fixed mass, the equivalent radius always produce the more anisotropic tidal field, whereas $\fRvir$ yields the more isotropic one. 

Interestingly, the median of $\log(q^2)$ obtained from setting $\Rs=\Req$ (black dotted curve) varies with protohalo mass exactly in the region where the change of slope in the colored curves occur. When filtered around $\Req$, the tidal field reaches a minimum level of anisotropy; smoothed below or above that value, the anisotropy is typically higher. Thus, this particular measure of the size of protohalos seems to define a characteristic scale for the correlation with the surrounding tidal field. The other two protohalo size measurements, $\Rlag$ and $\fRvir$, produce medians of $\log(q^2)$ that follow similar trends, but both are shifted toward lower anisotropies relative to the critical region where the color curves change slope. 

Finally, contrary to what we found for $\delta$, none of the protohalo size measurements produce a smoothed tidal field with a $q^2$ distribution that is independent of mass. When we consider their size as a filtering scale, protohalos of higher mass are typically embedded in more isotropic tidal fields, whether $\Rs$ is fixed to $\Req$, $\Rlag$, or $\fRvir$. For example, the median of $\log(q^2)$ corresponding to $\Rs = \Req$ changes from $-3.1\err{0.2}{0.2}$ for protohalos in the low-mass subsample, to $-3.3\err{0.2}{0.2}$ for intermediate-mass protohalos, and $-3.5\err{0.2}{0.2}$ for high-mass protohalos.

To further analyze the correlation between protohalos and their surrounding tidal field, in Figure \ref{fig:medalign_TFvsI} we explore their early alignment. Here we present the median of the cosine of the angle between the principal axes of the tidal field and the corresponding axes of the inertia tensor of protohalos, both measured in the initial conditions of the simulation (i.e., $z=80$). Unlike the previous figures, here we show only the medians as a function of protohalo mass, without the PDFs. The top row corresponds to results obtained by varying the filter scale from $\Rs = 0.5\hMpc$ to $26\hMpc$ for all halos, whereas the bottom row shows the alignment when different protohalo size-dependent filter lengths are applied. From left to right, each panel displays the median value of $|\ii\cdot\ti(\Rs)|$ for $k = 1, 2, 3$.

In the top row, the colored curves indicate that the alignment between $\mathbf{I}$ and $\mathbf{T}(\Rs)$ depends on both $M_\mathrm{FoF}$ and $\Rs$. At fixed mass, there is a characteristic scale at which the alignment reaches a maximum before decreasing. Conversely, when fixing the filter length, the variation of alignment differs for small and large $\Rs$: with increasing mass, $|\ii\cdot\ti(\Rs)|$ generally decreases for low $\Rs$ and increases for large $\Rs$. Consequently, there is a range of intermediate scales (approximately $1$ to $5\hMpc$) where the colored curves transition between these trends. This defines an envelope of maximum alignment, within which the orientations of the minor axes, $\ic$ and $\tc$ (top-right panel), exhibit the highest correlation of the three pairs.
 
  \begin{figure*}
	\includegraphics[width=1.8\columnwidth]{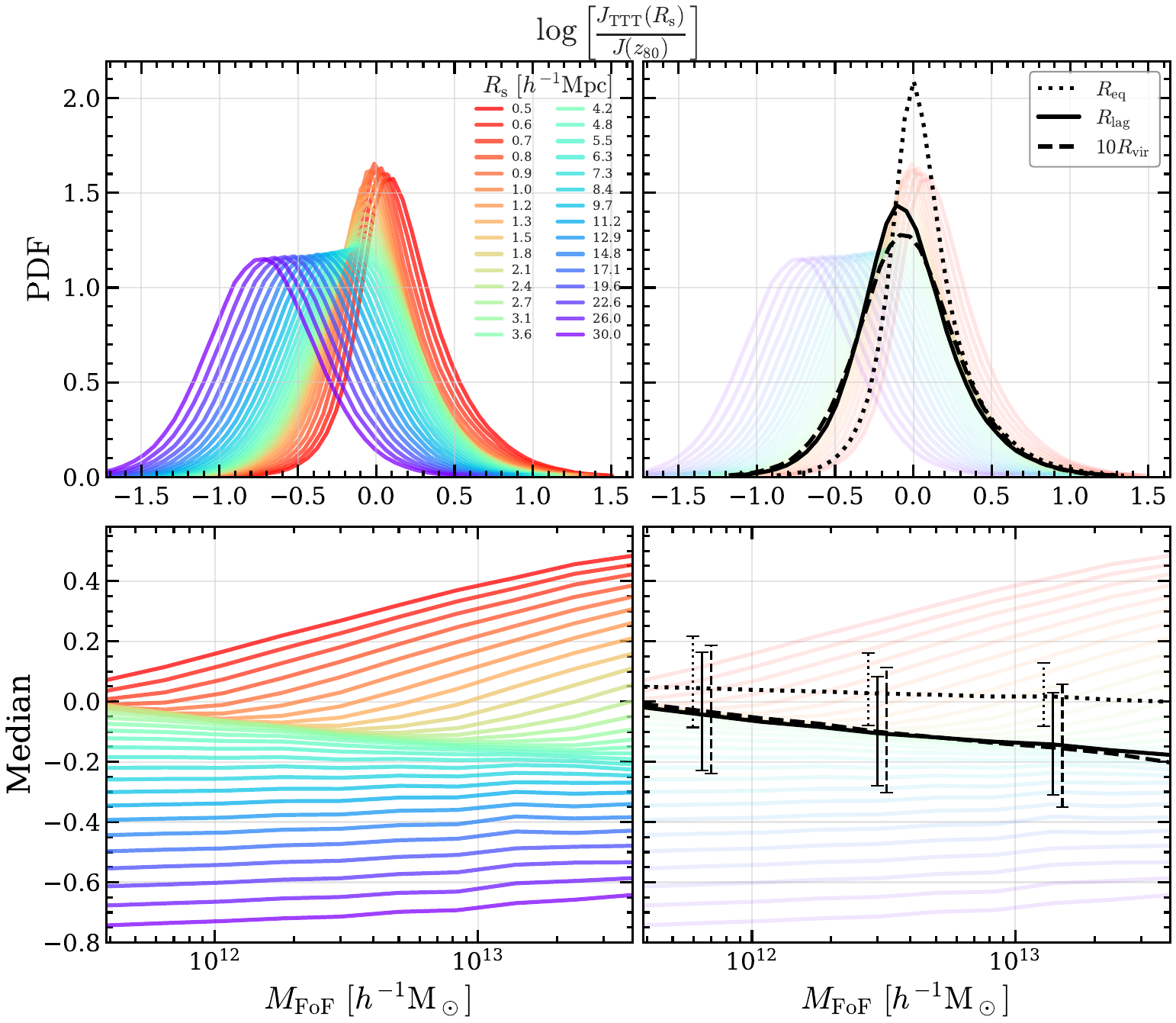}
   \caption{Ratio between the AM amplitude predicted by the TTT, $J_\mathrm{TTT}(\Rs)$, and the true AM of protohalos at $z=80$, $J(z_{80})$. The top row shows the PDFs of $\log\left[J_\mathrm{TTT}(\Rs)/J(z_{80})\right]$ for fixed smoothing scales $\Rs$ (left) and for size-dependent filters corresponding to $\Req$, $\Rlag$ and $10\Rvir$ (right). The bottom row shows the median values of the ratio as a function of protohalo mass, $M_\mathrm{FoF}$, for the same filter scales. Error bars in the bottom-right panel indicate the interquartile range for three different mass bins.}
   \label{fig:ratio_jTTT-jIC}
 \end{figure*}
 
Now we focus on the bottom row, i.e., the median alignment between $\mathbf{I}$ and $\mathbf{T}(\Rs)$ for $\Rs = \Req$, $\Rlag$, and $\fRvir$. We observe that the dependence on mass almost vanishes, although there is a slight tendency for high-mass protohalos to be less aligned with the surrounding tidal field, especially for $\Rs = \Req$. However, the most notable result is that the curve corresponding to $\Rs = \Rlag$ coincides almost exactly with the envelope of maximum alignment, for all three pairs of axes and for all masses. Hence, the Lagrangian radius appears to define a robust and consistent size-dependent scale that maximizes the correlation between $\mathbf{I}$ and $\mathbf{T}(\Rs)$, in agreement with \citet{leeetal2009}. 

Setting $\Rs = \Req$, on the other hand, produces a tidal tensor whose minor axis is strongly aligned with the minor axis of the corresponding inertia tensor but is considerably misaligned relative to the other two axes. This suggests that the tidal field around protohalos is more coherent along $\tc$ than in the other two directions.

In summary, the analysis of the properties of the tidal field at $z=80$ yields the following results:  
\begin{itemize}  
    \item For the density contrast, $\delta(\Rs)$, setting the smoothing scale of the tidal tensor to any protohalo size-dependent length produces a mass-independent distribution with a nearly constant median value. This median is highest when $\Rs = \Req$.  
    \item Regarding the tidal shear, $q^2(\Rs)$, for each protohalo mass there exists a particular smoothing scale where the anisotropy reaches a minimum. This characteristic scale approximately coincides with the equivalent radius, $\Req$, of each protohalo. Moreover, for any size-dependent filter length, the anisotropy decreases with increasing protohalo mass.  
    \item The alignment between the tidal field and the inertia tensor of each protohalo depends significantly on $\Rs$. For each protohalo mass, there is a characteristic smoothing scale that maximizes the alignment, which typically corresponds to the Lagrangian radius, $\Rlag$. The alignment is strongest along the direction of $\tc$, where the tidal field also appears more coherent.  
\end{itemize}

\section{Difference between true and predicted AM in the initial conditions}
\label{sec:TTT_predictions}

\begin{figure*}
    \includegraphics[width=1.8\columnwidth]{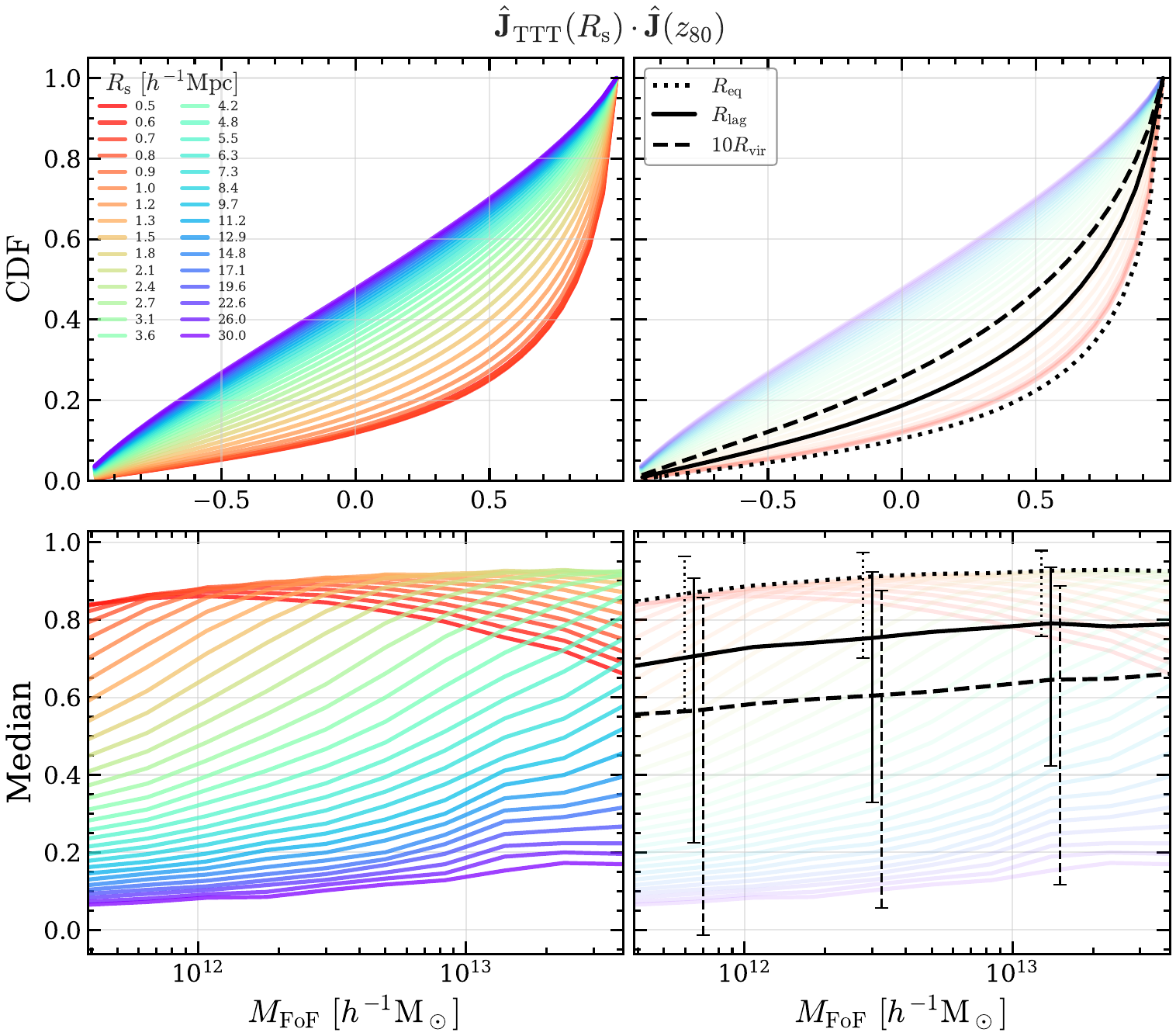}
    \caption{\emph{Left panel}: cummulative density function of the cosine of the angle between the predicted and the true direction of the AM at $z=80$, for different smoothing scales $\Rs$. \emph{Right panel}: median of the distributions as a function of protohalo mass.}
    \label{fig:align_jTTT-jIC}
\end{figure*}

We now assess the performance of the TTT in predicting the AM of protohalos at $z=80$, focusing on both its amplitude and orientation. To this end, we use Eq. \eqref{eq:TTT}, excluding the time-dependent factor, and compute the value that corresponds to each filter length $\Rs$: 
\begin{equation*}
J_{\mathrm{TTT},i}(\Rs) = \epsilon_{ijk} I_{jl} T_{lk}(\Rs),
\end{equation*}
appropriately normalized. We then compare these results with the actual AM of protohalos in the initial conditions, $\mathbf{J}(z_{80})$.

In the top-left panel of Figure \ref{fig:ratio_jTTT-jIC}, we show the ratio between the amplitude of the AM predicted by Eq. \eqref{eq:TTT}, $J_\mathrm{TTT} = |\mathbf{J}_\mathrm{TTT}|$, and the measured AM for each protohalo, $J(z_{80}) = |\mathbf{J}(z_{80})|$. The colored curves illustrate that as the smoothing scale $\Rs$ used to compute the tidal tensor increases, the AM predicted by the TTT typically decreases in amplitude. Starting from $\Rs = 0.5$, where predictions are slightly above $J(z_{80})$, the predicted AM systematically diminishes with increasing $\Rs$, quickly falling below the measured AM. Consequently, there exists a specific scale where the median of the ratio $J_\mathrm{TTT}/J(z_{80})$ is approximately 1.

From previous results, we expect this optimal scale to vary with protohalo mass. To explore this dependence, the bottom-left panel of Figure \ref{fig:ratio_jTTT-jIC} shows the median of $\log[J_\mathrm{TTT}(\Rs)/J(z_{80})]$ as a function of $M_\mathrm{FoF}$. For fixed $\Rs$ values above $\sim 3\hMpc$, the TTT underestimates the AM regardless of protohalo mass. For smaller smoothing scales, however, the predicted AM amplitude increases with protohalo mass. Each curve has a specific mass scale where the prediction is most accurate, and this mass increases with larger filter sizes. This trend reflects the relationship between protohalo mass and size: more massive protohalos correspond to larger Lagrangian regions, which are expected to couple with longer-wavelength modes of the tidal field.

Interestingly, this relationship can be approximately accounted for when setting $\Rs = \Req$, $\Rlag$, and $10\Rvir$. These results are shown in the right-hand panels of Figure \ref{fig:ratio_jTTT-jIC}. Surprisingly, using the Lagrangian radius $\Rlag$ does not yield the most accurate predictions when applying Eq. \eqref{eq:TTT}. Except for the lowest-mass protohalos, this scale tends to underestimate the AM amplitude, with the underestimation increasing for higher masses. For $\Rs = \Rlag$, the median of $\log[J_\mathrm{TTT}(\Rs)/J(z_{80})]$ at the positions of the error bars is $-0.04\err{0.17}{0.13}$, $-0.11\err{0.19}{0.17}$, and $-0.14\err{0.17}{0.17}$, from left to right. A similar trend with slightly greater scatter is observed for $\Rs = 10\Rvir$.

The protohalo size measure that yields the best prediction for the AM amplitude is the equivalent radius, $\Req$. The distribution for $\Rs = \Req$ in the top-right panel peaks near $J_\mathrm{TTT}(\Rs)/J(z_{80}) = 1$ and is narrower than those for the other two scales. Additionally, the variation of the median ratio with protohalo mass \hl{(bottom-right panel)} is less pronounced. Specifically, at the positions of the error bars, the medians are $0.04\err{0.17}{0.13}$, $0.03\err{0.13}{0.11}$, and $0.02\err{0.11}{0.10}$, from left to right.

Next, we evaluate the accuracy of the TTT in predicting the orientation of the AM. Figure \ref{fig:align_jTTT-jIC} shows the alignment between the predicted and true AM directions. The top-left panel presents the cumulative density function (CDF) of the cosine of the angle between the unit vectors $\Jdir_\mathrm{TTT}(\Rs) = \mathbf{J}_\mathrm{TTT}(\Rs)/J_\mathrm{TTT}(\Rs)$ and $\Jdir(z_{80}) = \mathbf{J}(z_{80})/J(z_{80})$, for fixed filter scales $\Rs$. 
The CDF (unlike the PDF) clearly distinguishes alignment trends by showing the cumulative probability of misalignment angles. A random distribution appears as a straight diagonal line, while curves bending below it indicate better-than-random alignment. We use the CDF here because the PDF would peak sharply near perfect alignment ($\cos\theta\sim1$), making differences between smoothing scales indistinguishable. Hence, we see that the alignment is generally stronger for small values of $\Rs$ and decreases monotonically as $\Rs$ increases. For very large smoothing scales (above $\sim 10 \hMpc$), the CDFs are nearly consistent with random orientations. 

The dependence of this result on protohalo mass is shown in the bottom-left panel, where we plot the median value of $\Jdir_\mathrm{TTT}(\Rs) \cdot \Jdir(z_{80})$ as a function of $M_\mathrm{FoF}$. Here, we observe that the predictions for low-mass protohalos are typically more accurate when $\Rs$ is small. In general, the model becomes less accurate as $\Rs$ increases. However, for higher-mass protohalos, the filter scale that provides the best alignment between the predicted and true AM directions shifts to larger values of $\Rs$.

\begin{figure*}
    \includegraphics[width=2\columnwidth]{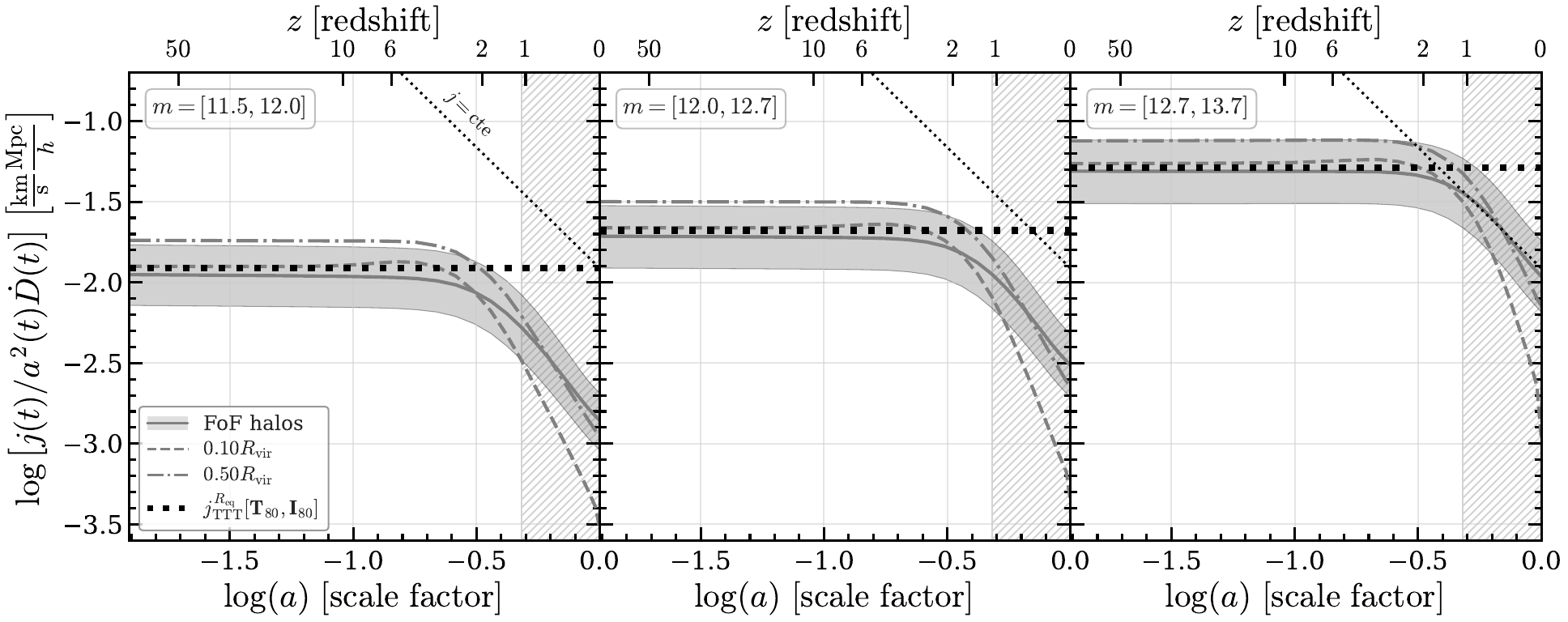}
    \caption{Evolution of the specific AM of protohalos, normalized to the expected growth factor predicted by the TTT, $a^2(t)\dot{D}(t)$. The solid gray curve in each panel represents the median evolution of $\log[j/a^2\dot{D}]$ for the full FoF halos, with the shaded region indicating the interquartile range. The dashed and dot-dashed gray curves show the same quantity but for particles that, at present time, reside within $0.1$ and $0.5$ of $\Rvir$, respectively. The thick dotted black line shows the prediction from the TTT (Eq. \ref{eq:TTT}), where both the inertia and tidal tensors are computed at $z=80$ using a smoothing scale equal to the equivalent radius, $\Req$, of each protohalo. The thin dotted black line in the upper right corner represents the slope associated with constant angular momentum. The shaded region at the right-hand side of each panel indicates the highly non-linear regime, where TTT predictions are not expected to hold. Each panel corresponds to protohalo samples of increasing mass, from left to right.}
    \label{fig:evol_J_amplitude}
\end{figure*}

In the right-hand panels of Figure \ref{fig:align_jTTT-jIC}, we repeat the alignment analysis, this time adjusting $\Rs$ based on each protohalo size. The CDFs in the top-right panel reveal that the equivalent radius ($\Req$, black dotted curve) provides the best prediction for the AM orientation. This result may seem unexpected, as the Lagrangian radius ($\Rlag$) is typically used as a reference scale for such analyses. However, this finding is consistent with the results for the AM amplitude. 

Regarding the dependence on protohalo mass, the bottom-right panel shows that for any fixed mass, the median alignment when $\Rs = \Req$ is always higher than for the other two size-dependent scales. Although there is a slight dependence on $M_\mathrm{FoF}$, with predictions for low-mass protohalos being less accurate than for high-mass ones, the black dotted curve closely follows the envelope of best alignments seen in the bottom-left panel. Consequently, the poor performance of the model for low masses may be attributed to a resolution limitation in our method's implementation.

Moreover, the equivalent radius not only produces better median alignments but also smaller interquartile ranges. Specifically, for the position of the error bars, $\Req$ yields median alignments of $0.87\err{0.09}{0.30}$, $0.91\err{0.06}{0.21}$, and $0.93\err{0.05}{0.17}$, from left to right, respectively. In contrast, $\Rlag$ produces medians of $0.71\err{0.20}{0.48}$, $0.75\err{0.17}{0.42}$, and $0.79\err{0.14}{0.37}$, while $10\Rvir$ results in $0.57\err{0.29}{0.58}$, $0.60\err{0.27}{0.55}$, and $0.65\err{0.24}{0.53}$. 

When $\Rs = \Req$, half of the low-mass protohalos have their predicted AM within $\sim30^\circ$ of the true direction, and $75\%$ are misaligned by less than $\sim56^\circ$\footnote{For random orientations, the expected misalignment is $90^\circ$.}. For high-mass protohalos, the predictions are even more accurate, with half being within $\sim22^\circ$ and $75\%$ within $\sim40^\circ$ of the true direction.

In summary, when implementing Eq. \eqref{eq:TTT} with a tidal tensor filtered at different scales, we observe the following:
\begin{itemize}
    \item The amplitude of the predicted AM varies significantly with both the smoothing scale $\Rs$ and protohalo mass $M_\mathrm{FoF}$. Among the three size measures we tested ($\Req$, $\Rlag$ and $\fRvir$, the most accurate prediction for the AM amplitude is obtained when the smoothing scale is set to the equivalent radius of each protohalo, $\Req$.
    \item The true direction of the AM is also best predicted when $\Rs = \Req$. This holds for the overall sample as well as for subsamples across the entire mass range.
\end{itemize}

\section{Deviations in the evolution of the AM}
\label{sec:spin_evolution}

Having identified the optimal scale for implementing the TTT and predicting the AM of protohalos at $z=80$, we now turn to investigate how its evolution deviates from the model's expectations. Our goal is to explore ways to characterize and, if possible, account for these deviations. To achieve this, we analyze the true AM of protohalos at various redshifts and study its dependence on the regions of the virialized halos that the particles occupy at $z=0$. Additionally, we re-implement Eq. \eqref{eq:TTT}, measuring both the inertia tensor and the tidal field at different stages of the protohalos' linear and quasi-linear evolution, to understand whether their variation can explain the observed deviations from the model.

\subsection{Amplitude}
\label{sec:spin_evolution:amplitude}

\begin{figure*}[H]
    \includegraphics[width=2\columnwidth]{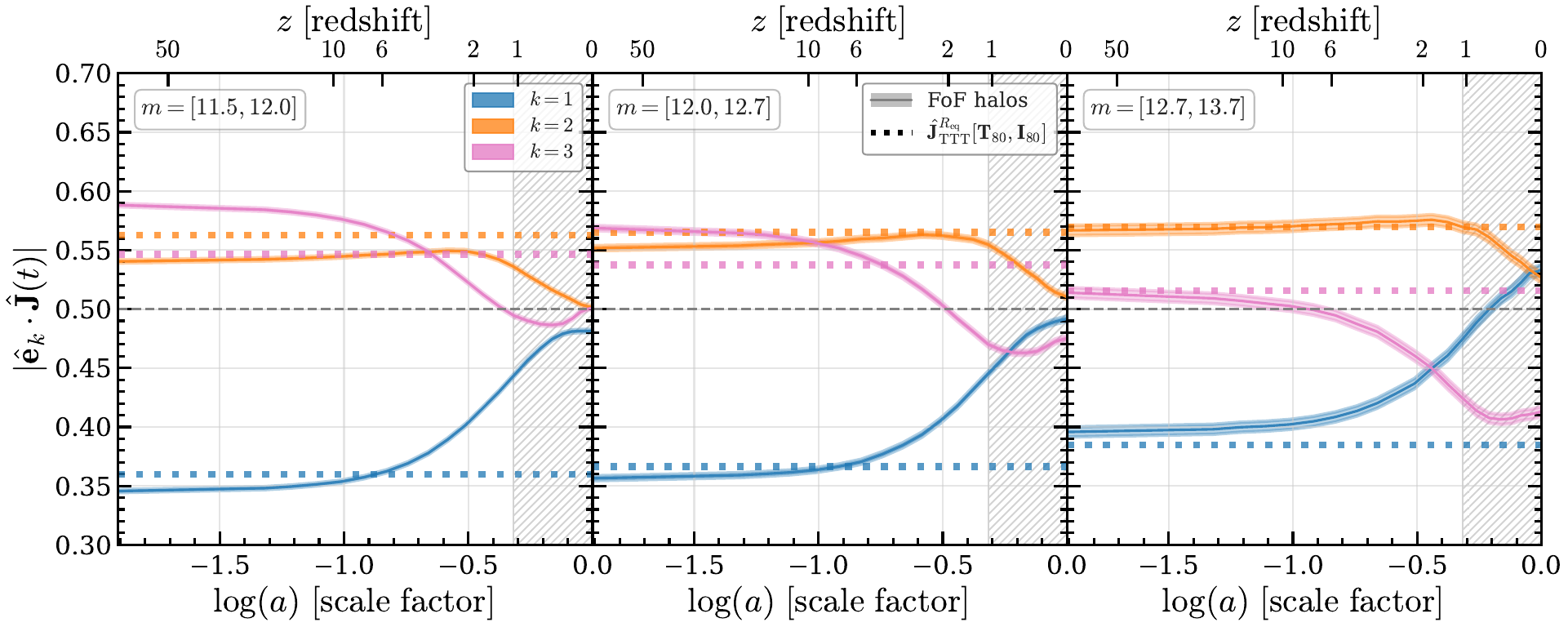}
    \caption{Evolution of the median alignment between the AM direction, $\Jdir(t)$, and the principal axes of the present-day filaments. The alignment is quantified as $\left|\ei \cdot \Jdir(t)\right|$, where $k=1,2,3$ correspond to the first axis of collapse ($\ea$, blue curves), the intermediate axis ($\eb$, orange curves), and the spine of the filaments ($\ec$, pink curves), respectively. The shaded region  around each of these curves indicates the $3\sigma$ error in the computation of the median. The dotted horizontal lines represent the TTT predictions at $z=80$, assuming a smoothing scale $\Rs = \Req$. The shaded region at the right-hand side of each panel indicates the highly non-linear regime, where TTT predictions are not expected to hold. Each panel corresponds to protohalo samples of increasing mass, from left to right.}
    \label{fig:evol_J-F_align}
\end{figure*}

In Figure \ref{fig:evol_J_amplitude} we show the median evolution of the specific AM amplitude, both measured in the simulation and predicted by our implementation of the TTT. The gray solid lines represent the median of the AM distribution of protohalos at each snapshot, whereas the surrounding gray shaded regions indicate the corresponding interquartile range. The thick black dotted lines show the expected behavior according to Eq. \eqref{eq:TTT}, with a smoothing scale $\Rs=\Req$ for each protohalo. From left to right, each panel corresponds to a subsample of protohalos of increasing mass, as indicated by the top-left labels. The shaded region in the righ-hand side of each panel indicates the final, highly non-linear stage, where the TTT is not expected to be valid.

The first thing we confirm is that the TTT works very well during the early stages of structure formation. Quantitatively, we observe a slight overestimation of the AM when $\Rs = \Req$, primarily for low-mass systems. Qualitatively, the model describes the evolution very well up to $z \sim 4$ for low-mass protohalos and up to $z \sim 2$ for more massive protohalos. It is only then that the true evolution significantly deviates from the expected evolution (horizontal line).

To better understand how the assembly of present-day halos influences this process, we analyze the specific AM of particles that, at $z = 0$, are located within spheres of radii $0.1\Rvir$ and $0.5\Rvir$. The evolution of these particles is shown in Figure \ref{fig:evol_J_amplitude} by the gray dashed and dot-dashed curves, respectively.

As expected, particles residing in the inner regions of the halo at present time exhibit lower AM at late times compared to those in the outer regions. This trend becomes particularly evident during the final stages of evolution, where the specific AM of these particles decreases sharply. \hl{As illustrated by the thin dotted black curve, which represents the case of constant AM in these diagrams, this sharp decline indicates a net loss of AM as the particles settle into the inner regions of the protohalos.}

Interestingly, the earlier stages of evolution reveal a more complex picture. Despite their current location in the inner halo, these particles initially had comparable or even higher specific AM than those located in the outer regions of the protohalo. \hl{In effect, Figure \ref{fig:evol_J_amplitude} shows that, from $z=80$ down to $z\sim 2$, the median specific AM of the particles that end up at $0.5\Rvir$ at $z=0$ (and, to a lesser extent, also those at $0.1\Rvir$) is higher than that of the total FOF halo, once all values are normalized by $a^2\dot{D}$. In the regime where both curves are roughly constant, one clearly lies above the other. For example, for low-mass halos, the median $\log(j/a^2\dot{D})$ for the $0.5\Rvir$ particles is $\sim -1.75$, whereas for the total FOF halo it is $\sim -1.95$. In linear terms, this implies that the typical specific AM of the inner particles is about $60\%$ higher than that of the full FOF protohalo.}

This result suggests that particles contributing to the protohalo span a broad range of specific AM values and that significant redistribution of AM occurs throughout the halo's assembly process. By the end of the simulation, particles that lose a substantial fraction of their specific AM to the outer regions tend to migrate inward, settling into the halo's core.

\subsection{Direction}
\label{sec:spin_evolution:direction}
 
To investigate the evolution of the AM orientation, we compare the AM direction of protohalos at different times with fixed reference directions. Specifically, we focus on the principal axes of the closest present-day filaments in the cosmic web, as identified using the \nexus{} method \citep{cautunetal2013}. These filaments provide a meaningful reference for understanding alignments between AM and the large-scale structure, a topic of interest in recent studies \citep[e.g.][]{tempelylibeskind2013,zhangetal2015,welkeretal2019,kraljicetal2021,desaiyryden2022}. 

The \nexus{} framework is particularly suited for this analysis because it uses a multiscale approach to detect structures with varying degrees of prominence around halos, independent of their size or shape. This scale-free property makes it an excellent tool for comparing the preferred directions of present-day filaments with the early tidal field at $z=80$, computed using smoothing scales tied to protohalo sizes. The alignment between the AM and the closest filament thus serves as both an observational reference and a proxy for understanding how the early tidal field influences protohalo AM, as explored in previous studies \citep{lopezetal2021}.

\begin{figure*}[H]
    \includegraphics[width=2\columnwidth]{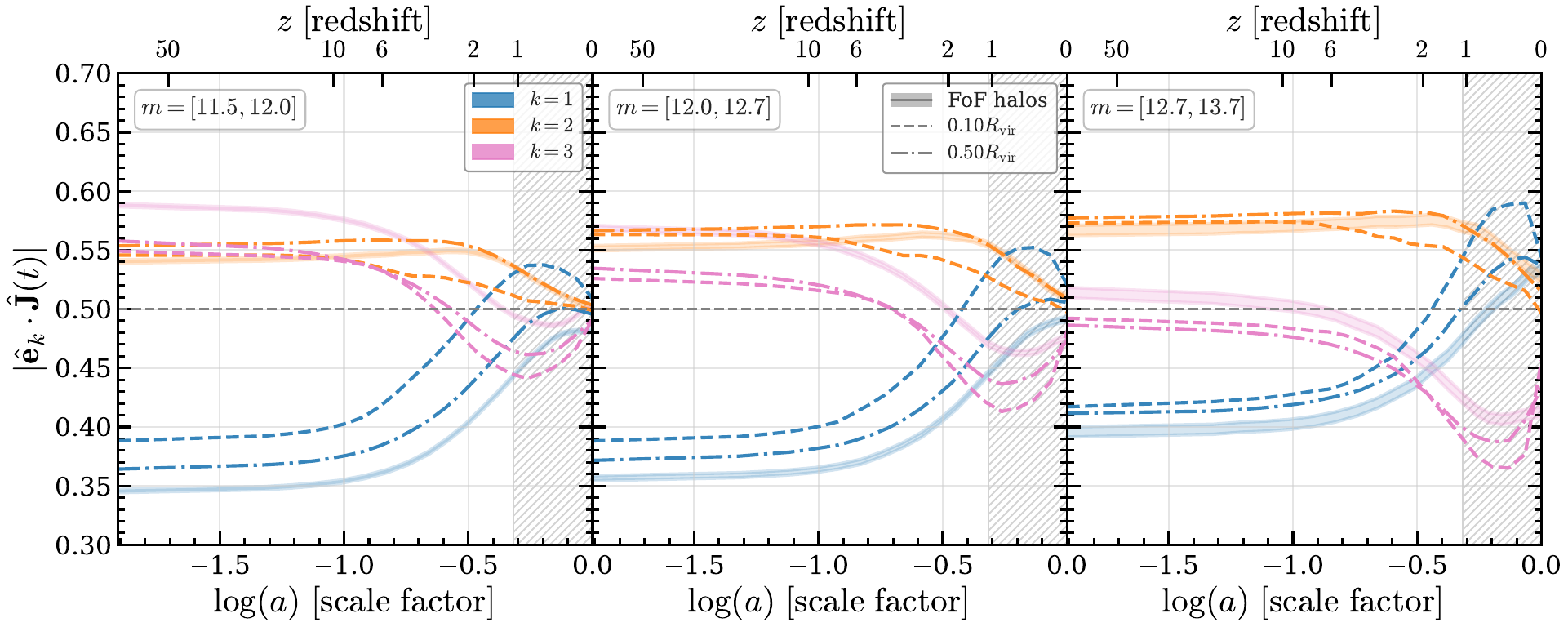}
    \caption{Evolution of the median alignment between the inner AM of protohalos and the principal axes of the present-day filaments. The dashed and dot-dashed curves show the median value of $\left|\ei \cdot \Jdir(t)\right|$ for particles that, at $z=0$, reside within $0.1$ and $0.5$ of $\Rvir$, respectively. For comparison, the light shaded curves show the alignment for the full FoF protohalos. The shaded region at the right-hand side of each panel indicates the highly non-linear regime, where TTT predictions are not expected to hold. Each panel corresponds to protohalo samples of increasing mass, from left to right.}
    \label{fig:evol_Jvir-F_align}
\end{figure*}
 
In Figure \ref{fig:evol_J-F_align}, we present the evolution of the median alignment between the true AM direction and the main axes of the filaments at present time. The thick blue, orange, and pink curves represent the median of $|\ei \cdot \Jdir(t)|$, for $k=1, 2, 3$, respectively. Here, both $\ei$ and $\Jdir(t)$ represent unit vectors. Each panel corresponds to different subsamples of protohalos, with increasing mass from left to right. For comparison, the dotted lines show the prediction of the TTT using Eq. \eqref{eq:TTT} at $z=80$, where the smoothing scale is set to the equivalent radius of each protohalo, $\Rs = \Req$. Within the framework of the standard TTT implementation, the AM direction remains constant during the linear and quasi-linear regimes, and thus the predictions are represented as horizontal lines. The shaded region on the right-hand side of each panel indicates the highly non-linear stage, where the TTT is no longer expected to hold.

We first observe that the alignment between the true AM direction and the filament axes evolves significantly from the initial conditions of the simulation to the present time. Specifically, the orientation of $\Jdir$ with respect to $\ec$, which corresponds to the spine of the filaments, transitions from being preferentially aligned at $z=80$ to predominantly perpendicular at $z=0$. However, this evolution is clearly mass-dependent. Low-mass protohalos initially exhibit a strong alignment with $\ec$. As redshift decreases, this alignment weakens, becoming preferentially perpendicular around $z \sim 1$ (i.e., $|\ec \cdot \Jdir(t)| < 0.5$). In the highly non-linear regime, this trend reverses, and low-mass protohalos reach $z=0$ with orientations that can be considered random. Higher-mass protohalos follow a similar trend but start with weaker alignment, cross the $0.5$ threshold earlier, and at $z=0$ show a stronger tendency toward perpendicularity compared to their low-mass counterparts.

The evolution of $\Jdir$ with respect to $\ea$, the first axis of collapse, exhibits distinct behavior. At $z=80$, the AM direction is initially strongly perpendicular to $\ea$. As redshift decreases, the alignment steadily increases, but low- and medium-mass protohalos never exceed the $0.5$ threshold. For the most massive protohalos, however, the alignment continues to grow after $z \sim 1$, and by $z=0$, their AM is weakly aligned with $\ea$.

\begin{figure*}
    \includegraphics[width=2\columnwidth]{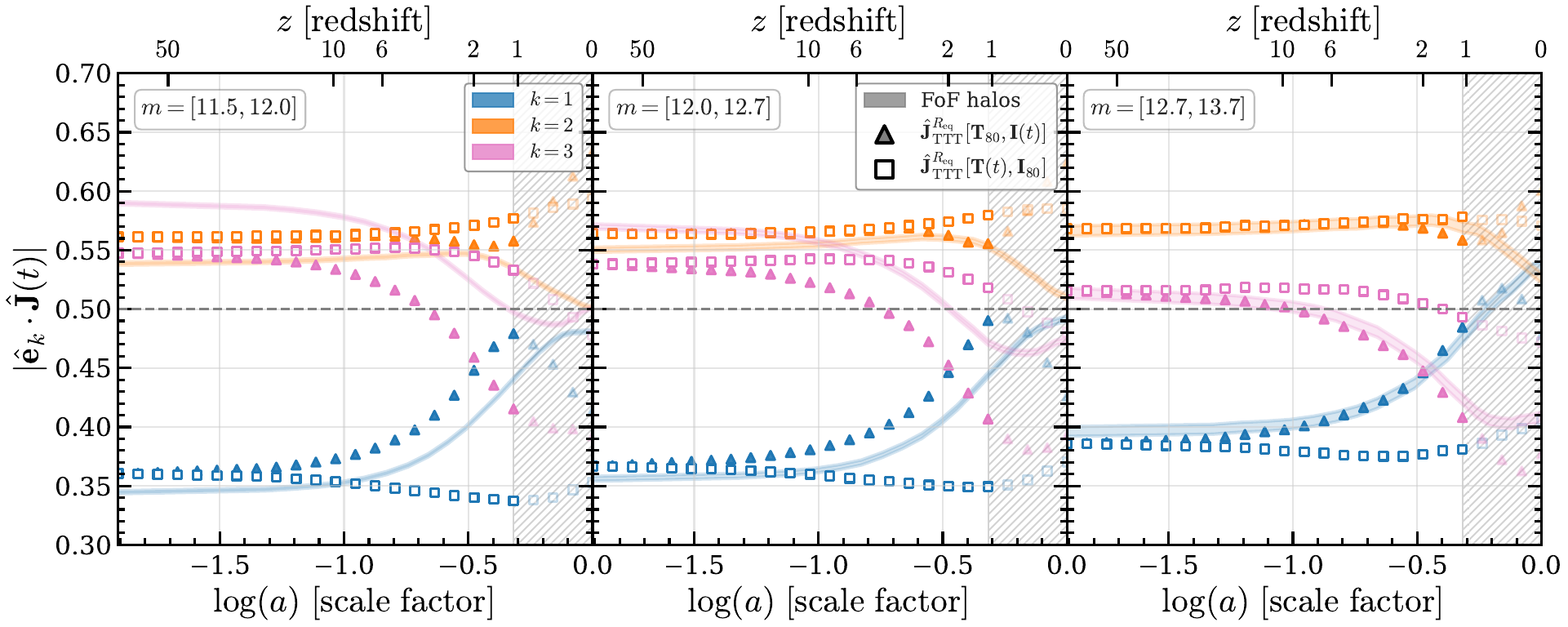}
    \caption{Evolution of the median alignment between the AM predicted by TTT at each time step and the principal axes of the present-day filaments. The filled triangles show the prediction of the TTT when we fix the tidal tensor at its $z=80$ value, $\mathbf{T}_{80}$, but allow the inertia tensor $\mathbf{I}(t)$ to change in time. Conversely, the empty squares represent the prediction when fixing the inertia tensor at the initial conditions, $\mathbf{I}_{80}$, and allowing the tidal tensor to evolve, $\mathbf{T}(t)$. For comparison, the light shaded curves show the alignment for the full FoF protohalos. The shaded region at the right-hand side of each panel indicates the highly non-linear regime, where TTT predictions are not expected to hold. Each panel corresponds to protohalo samples of increasing mass, from left to right.}
    \label{fig:evol_J-F_align_TF0-IT0}
\end{figure*}
 
Finally, the alignment of $\Jdir$ with $\eb$, the intermediate axis of the filament, begins with a preference toward alignment. This tendency is more pronounced for higher-mass protohalos. Unlike the trends observed with the other two axes, the median of $|\eb \cdot \Jdir(t)|$ remains approximately constant during intermediate redshifts. A noticeable shift occurs only after $z \sim 1$, when protohalos enter the highly non-linear regime. At this stage, the alignment decreases, and by $z=0$, low-mass protohalos show random orientations, while higher-mass protohalos retain a marginal alignment with $\eb$.

When comparing the true evolution of the spin-filament alignment to the predictions of our TTT implementation, we find, as expected, better agreement at high redshift than at present time. However, significant deviations from the TTT predictions appear earlier than typically assumed, specifically between $z \sim 10$ and $z \sim 1$. During this period, the median alignment of the AM direction with $\eb$ (the intermediate axis of the filaments) remains nearly constant, but its orientation with respect to $\ea$ (the first collapse axis) and $\ec$ (the filament spine) evolves systematically. More precisely, we observe a consistent rotation of the AM around $\eb$, which leads to a shift in alignment from $\ec$ toward $\ea$.

The accumulated effect of this systematic evolution depends on the initial state of the protohalos and reveals a clear mass dependence. Low-mass protohalos are initially strongly aligned with $\ec$. As the AM rotates around $\eb$, the alignment with $\ec$ weakens over time. By $z \sim 1$, this decay produces nearly random configurations of the AM direction with respect to the filament spine, while the alignment with $\eb$ remains almost unchanged. In contrast, high-mass protohalos start with weaker alignment to $\ec$. As this alignment diminishes, a preference for perpendicular configurations relative to the filament spine develops by $z \sim 1$. 

In the highly non-linear regime, spanning from $z \sim 1$ to the present time, the AM direction generally undergoes significant, seemingly chaotic changes. This behavior is evidenced by the median alignment curves approaching $0.5$, indicative of random configurations. During this stage, for example, the previously stable alignment with respect to $\eb$ diminishes as the AM orientation is increasingly randomized. While this late evolution tends to wash out or stall the systematic trends established in earlier stages, it does not appear to be the primary driver of the present-day AM configuration. Instead, the results suggest that the $z=0$ alignments are largely shaped by the preceding quasi-linear evolution. The only consistent trend observed during the highly non-linear regime is the increasing alignment of the AM of high-mass protohalos with $\ea$, indicating that the non-linear interactions may act to reinforce the earlier evolution in these cases.

We now turn to analyze the AM direction of particles that, at $z = 0$, are located within spheres of radii $0.1\Rvir$ and $0.5\Rvir$. The evolution of these particles is shown in Figure \ref{fig:evol_Jvir-F_align} by the dashed and dot-dashed curves, respectively. For comparison, we show with light shaded colored curves the results for the full FoF protohalos. Interestingly, the particles located deeper inside the present-day halos do not retain their initial AM orientation better than those found in the outer regions, as one might expect. Instead, these particles follow the same overall trends observed for the full protohalos, and in some cases, they show an even more pronounced evolution, particularly with respect to $\ea$ and $\ec$.  

At $z\sim1$, the AM direction of particles within $0.1\Rvir$ shows a clear ``flip'' with respect to $\ea$ for all masses, in the sense that their median alignment has already surpassed the $0.5$ threshold, despite starting from values below this threshold. At the same time, the alignment with respect to $\ec$ becomes preferentially perpendicular. These trends are qualitatively similar to those seen for the full protohalos but are more pronounced for the deepest particles. During the highly non-linear regime ($z<1$), the AM of these particles exhibits a stronger tendency to randomize. This effect is reflected by the median alignment values converging more sharply to $0.5$, particularly with respect to $\ea$ and $\ec$, at the right-hand side of each panel. The closer the particles are to the center of the halo at $z=0$, the stronger their alignment trends are ``washed out'' by non-linear interactions.  

\subsection{Mass dependent Spin-Flip}
An additional key result is that, at present time, the AM direction of the innermost particles does not display the well-known mass-dependent ``spin flip'' trend observed in full halos \citep{aragoncalvoetal2007,hahnetal2007,codisetal2015,veenaetal2018,veenaetal2021}. For the total halo spin, it is typically found that low-mass halos show an alignment with $\ec$, the filament spine, while high-mass halos tend to exhibit perpendicular AM alignments. In contrast, for particles at $0.1\Rvir$, we find that the AM is perpendicular to $\ec$ regardless of the halo mass. 

However, the origin of this behavior is not as previously thought. Earlier interpretations suggested that the spin flip arises due to late-accreted material, which primarily resides in the outer regions of halos at $z=0$, while the particles deeper inside the halo were believed to retain their original alignment. This explanation posited that environmental effects, such as vorticity within filaments or preferred directions of secondary accretion, dominate the spin flip. In contrast, our results indicate that particles at different depths inside the halo ($0.1\Rvir$ vs $0.5\Rvir$) experience similar evolution histories. The main difference lies in their initial alignment states and in their late evolution. 

In effect, particles located deeper inside the halo at $z=0$ start off from less aligned configurations, with median alignment values closer to $0.5$. As a result, the quasi-linear evolution that systematically shifts the AM orientation impacts these particles more strongly: for all masses, the AM tends to align with $\ea$ and become perpendicular to $\ec$. Hence, at $z\sim1$, inner particles show systematically stronger alignment signals with the filaments than what we observe for the full protohalos.

This enhanced alignment, however, does not persist. During the highly non-linear stages, the angular momentum of inner particles undergoes significant reorientations, diluting the imprint of the earlier tidal torques. These late-time variations appear largely uncorrelated with the surrounding filamentary structure, effectively erasing any previously established alignment or perpendicularity with respect to the cosmic web. This transition coincides with the onset of a clumpy mass distribution within halos (see Fig.~1), hinting at the role of hierarchical assembly. Our results are consistent with those of \citet{bettyfrenk2012} and \citet{bettyfrenk2016}, who show that inner particles are most affected during mergers. Such events likely bring together substructures that have acquired angular momentum from different tidal fields, transferring this orbital momentum to the spin of the remnant system.

\subsection{Nonlinear TTT transition}
\label{sec:spin_evolution:evolving_I-T}
To assess the extent to which the variations of the inertia tensor, $\mathbf{I}$, and the tidal tensor, $\mathbf{T}$, during the linear and quasi-linear regimes affect the predictions of the TTT, we compute $\mathbf{I}(t)$ and $\mathbf{T}(t)$ at multiple redshifts for the protohalos in our sample. For $\mathbf{I}(t)$, we use the positions of the protohalo particles, and for $\mathbf{T}(t)$, we follow the same procedure as in the initial conditions, applying the protohalo size-dependent smoothing scale, $\Rs = \Req$. Using Eq. \eqref{eq:TTT}, we generate two sets of predictions: one with a fixed tidal tensor and time-varying inertia tensor, $\Jdir_\mathrm{TTT}^{\Rs}[\mathbf{T}_{80}, \mathbf{I}(t)]$, and the other with a fixed inertia tensor and time-varying tidal tensor, $\Jdir_\mathrm{TTT}^{\Rs}[\mathbf{T}(t), \mathbf{I}_{80}]$. These are represented in Figure \ref{fig:evol_J-F_align_TF0-IT0} by filled triangles and empty squares, respectively. For comparison, the light shaded curves show the true evolution of the AM direction of the protohalos.

Both sets of predictions begin to diverge from the true AM direction at approximately the same redshift, $z\sim10$. This is also when the true AM starts to deviate from the fixed $z=80$ TTT predictions. This suggests that this moment in cosmic history is pivotal, marking the onset of deviations driven by processes not accounted for in the standard TTT implementation. Notably, the predictions with fixed $\mathbf{T}_{80}$ and varying $\mathbf{I}(t)$ qualitatively reproduce the trend of the true AM direction up to $z \sim 1$. 

In effect, for low-mass protohalos, the filled triangles (fixed $\mathbf{T}_{80}$, varying $\mathbf{I}(t)$) deviate slightly from the true AM direction, with the offset emerging in the initial conditions. This discrepancy is likely due to inaccuracies in the computation of $\mathbf{T}_{80}$, as we already noted in the previous section. Despite this limitation, the systematic trends seen in the true evolution, such as the decrease in alignment with $\ec$ and the increase in alignment with $\ea$, are reasonably well captured by the predictions when the inertia tensor evolves. The alignment with $\eb$ remains nearly constant throughout this period. For high-mass protohalos, the predictions of $\Jdir_\mathrm{TTT}^{\Rs}[\mathbf{T}_{80}, \mathbf{I}(t)]$ are excellent, closely matching the true median alignment up to $z\sim1$. This indicates that the model can effectively capture the shift in AM orientation of high-mass protohalos when the variation of the inertia tensor is taken into account. 

In contrast, the predictions of $\Jdir_\mathrm{TTT}^{\Rs}[\mathbf{T}(t), \mathbf{I}_{80}]$, which allow $\mathbf{T}(t)$ to vary while keeping $\mathbf{I}_{80}$ fixed, fail to reproduce the true evolution of the AM direction. These predictions deviate significantly from the light shaded curves as early as $z\sim10$. The alignment with $\eb$ and $\ec$ remains nearly constant, while the alignment with $\ea$ actually decreases. This behavior demonstrates that the variation of the tidal tensor over time does not play a significant role in driving the deviations observed in the true AM evolution. 

These results strongly suggest that the deviations of the true AM direction from the TTT predictions between $z\sim10$ and $z\sim1$ arise primarily from the evolution of the shape of the protohalos, as captured by their inertia tensor. Thus, modeling the evolution of $\mathbf{I}(t)$ during the linear and quasi-linear regimes might be sufficient to account for these deviations. In contrast, the time variation of the tidal tensor, $\mathbf{T}(t)$, does not significantly affect the accuracy of the predictions. 

\section{Summary and conclusions}
\label{sec:summary}

In this work, we have investigated how the tidal field and the characteristic scale of protohalos influence the origin and evolution of their AM. Using a dark matter-only cosmological simulation, we followed the formation and growth of protohalos from the initial conditions at $z = 80$ to the present day. We paid particular attention to the role of filtering scale in shaping the local tidal field and in determining the predictions of the TTT at early times. By analyzing the time evolution of the AM amplitude and its alignment with the present-day filamentary structure, we identified the stages at which the true AM trajectory departs from TTT expectations.

We have shown that the local overdensity around protohalos, $\delta$, and the tidal shear, $q^2$, both exhibit strong dependencies on the smoothing scale used to compute the tidal field. At fixed smoothing length, more massive protohalos tend to reside in denser and more isotropic environments. However, when the smoothing scale is adapted to the size of each protohalo, particularly using the equivalent radius $\Req$ (which is roughly half of the Lagrangian radius of protohalos and approximately three times the virial radius of present-day halos), the overdensity becomes nearly independent of mass.

In contrast, the tidal anisotropy retains a residual dependence on mass even when filtered at protohalo-specific scales. The equivalent radius also marks a transitional scale in the $q^2$–mass relation, where the tidal field reaches a minimum anisotropy. These results indicate that $\Req$ captures a physically meaningful boundary between the inner and outer tidal environments. While overdensities appear to reflect local conditions set at early times, tidal anisotropies preserve a connection with larger-scale structure and halo mass.

We have evaluated the ability of the TTT to predict both the amplitude and direction of protohalo AM at high redshift. Our results show that the accuracy of TTT predictions depends sensitively on the smoothing scale used to compute the tidal field. At fixed smoothing lengths, smaller scales yield more accurate predictions for both the amplitude and orientation of the AM, particularly for low-mass protohalos. However, the optimal filter scale increases with protohalo mass, reflecting the coupling of larger protohalos to longer-wavelength tidal modes. \hl{Crucially, among the three protohalo size measures tested ($\Req$, $\Rlag$ and $\fRvir$), the model achieves its best overall performance when the the filter scale is adjusted to the equivalent radius $\Req$ of each protohalo}.

Specifically, $\Req$ yields AM amplitude predictions with median ratios close to unity and the narrowest distributions, across all masses. It also results in the highest alignment between the predicted and true AM directions, with significantly smaller scatter compared to using the Lagrangian radius or virial-based scales. These findings, once again, indicate that $\Req$ marks a physically meaningful scale where the coupling between protohalo shape and the tidal field is optimally captured. The relatively poor performance of TTT for low-mass protohalos may be partly attributed to resolution limitations, but overall, our results highlight the importance of using adaptive, physically motivated smoothing scales in applications of TTT.

The evolution of AM in protohalos reveals a complex interplay between early linear growth, quasi-linear reconfigurations, and late-time non-linear dynamics. Although our analysis confirms that the TTT accurately predicts the amplitude of the AM in the early universe, deviations from the model arise as early as $z \sim 4$ for low-mass systems and $z \sim 2$ for higher-mass ones. These deviations correlate with the redistribution of AM during halo assembly: particles that ultimately settle into the inner regions of present-day halos often start with high specific AM, which they gradually lose over time. This redistribution contributes to the eventual decoupling between the initial AM configuration and its late-time state.

In terms of direction, our results demonstrate a systematic and mass-dependent evolution of the AM orientation relative to the cosmic web. Early alignments with the spine of present-day filaments ($\ec$) degrade over time, particularly due to a gradual rotation of the AM vector around the intermediate filament axis ($\eb$), effectively transferring alignment toward the principal axis of collapse ($\ea$). \hl{This transition is most evident during the quasi-linear regime ($z \sim 10$ to $z \sim 1$), after which the highly non-linear regime tends to further degrade AM coherence. In this stage, the evolution can involve genuinely chaotic trajectories in some systems (e.g. due to the interaction of multiple halos), but in other cases it may follow more predictable, non-linear dynamics that are not captured by TTT. Despite this degradation, some coherence persists, especially for massive halos. Notably, the final alignments at $z = 0$ appear more strongly shaped by the quasi-linear evolution than by the later transformations of the highly non-linear stage.} These findings suggest that the breakdown of TTT is not abrupt but unfolds progressively, driven by the combined effect of internal dynamical processes, the differential coupling with the large-scale tidal field, and the evolving structure of the cosmic web.

We also find that the AM direction of particles located deep inside present-day halos does not preserve its initial orientation better than that of outer particles. Instead, their evolution closely follows the trends seen for full protohalos and even shows stronger changes. This is particularly evident for alignments with the filament directions $\ea$ and $\ec$, which evolve more rapidly for particles ending up near the halo center.


More specifically, particles located within $0.1\Rvir$ at $z=0$ begin with weaker alignments and are more strongly affected by the quasi-linear evolution that reorients their AM toward $\ea$ and away from $\ec$. At $z \sim 1$, they show enhanced alignment signals compared to the full protohalos. However, this alignment is subsequently erased during the highly nonlinear regime, as these particles experience stronger reorientations, likely due to AM redistribution and the clumpy assembly of halos. These late-time changes typically decouple the AM direction from the cosmic web.

To identify the source of early deviations from the TTT, we tested variants of the model that allow for time evolution in either the inertia tensor or the tidal field. We find that the shape evolution of the protohalos, encoded in the time-varying inertia tensor $\mathbf{I}(t)$, is the main driver of the true AM evolution between $z \sim 10$ and $z \sim 1$. In contrast, the variation of the tidal tensor $\mathbf{T}(t)$ plays a negligible role. Predictions that include evolving $\mathbf{I}(t)$ match the trends followed by the true AM direction well, especially for high-mass halos, while those based on evolving $\mathbf{T}(t)$ fail early on.

\hl{Altogether, our results show that both the depth of particles within halos and the time evolution of the protohalo shape critically influence the AM direction. The evolution from the initial conditions, as predicted by the standard TTT, to the stage we refer to as ``highly non-linear'' is not purely stochastic. Instead, it shows a strong correlation with the direction of the cosmic web, and in the quasi-linear regime this evolution appears to be statistically predictable when the time dependence of the inertia tensor is considered, as shown in Section \ref{sec:spin_evolution:evolving_I-T}. This predictable component suggests that part of the deviations from standard TTT can be understood in terms of systematic shape changes rather than random, late-time processes.}

\hl{In this picture, early quasi-linear torques can even generate strong alignments even for initially weakly aligned systems (as we see, for instance, in Figure \ref{fig:evol_J-F_align}, between $\Jdir$ and $\ec$ for high-mass halos). Although the final stages of evolution can partially dilute these alignments, the $z=0$ signal still appears to reflect more strongly the imprint of the quasi-linear evolution than the influence of highly non-linear effects. The fact that a substantial fraction of the alignment evolution occurs in a regime where shape changes are measurable and predictable opens the possibility of improving analytical models without the need to explicitly incorporate all the details of highly non-linear dynamics. Properly modeling the inertia tensor evolution during this phase is thus essential to extend the predictive range of TTT beyond the initial linear regime.}

\hl{In this context, the modified TTT formulation proposed by \citet{lopez2023} provides an interesting and complementary perspective. Their approach introduces anisotropic scale factors as a proxy for the influence of the large-scale tidal field, which in practice leads to systematic and predictable changes in halo shapes. While their model does not explicitly track the time evolution of the inertia tensor, it may nevertheless capture some of the same physical effects identified in our analysis, namely, that controlled and measurable shape variations can play a decisive role in setting the final ($z=0$) AM alignments. This connection suggests that further exploration of shape-driven extensions to TTT could provide a practical path toward more accurate and physically motivated predictions for halo AM correlations.}

\section*{Acknowledgements}
This work has been partially supported by the Argentine \emph{Consejo Nacional de Investigaciones Científicas y Técnicas} (CONICET). This work used computational resources from UNC Supercómputo (CCAD) – Universidad Nacional de Córdoba (https://supercomputo.unc.edu.ar), which are part of SNCAD, República Argentina. \hl{We also thank the referee Mark Neyrinck for his insightful comments, which have helped to significantly improve this paper.}

\section*{Data Availability}
The data used in this work is available upon reasonable request to the corresponding author.





\bibliographystyle{mnras}
\bibliography{bibliografia} 

\begin{thebibliography}{}
\makeatletter
\relax
\def\mn@urlcharsother{\let\do\@makeother \do\$\do\&\do\#\do\^\do\_\do\%\do\~}
\def\mn@doi{\begingroup\mn@urlcharsother \@ifnextchar [ {\mn@doi@}
  {\mn@doi@[]}}
\def\mn@doi@[#1]#2{\def\@tempa{#1}\ifx\@tempa\@empty \href
  {http://dx.doi.org/#2} {doi:#2}\else \href {http://dx.doi.org/#2} {#1}\fi
  \endgroup}
\def\mn@eprint#1#2{\mn@eprint@#1:#2::\@nil}
\def\mn@eprint@arXiv#1{\href {http://arxiv.org/abs/#1} {{\tt arXiv:#1}}}
\def\mn@eprint@dblp#1{\href {http://dblp.uni-trier.de/rec/bibtex/#1.xml}
  {dblp:#1}}
\def\mn@eprint@#1:#2:#3:#4\@nil{\def\@tempa {#1}\def\@tempb {#2}\def\@tempc
  {#3}\ifx \@tempc \@empty \let \@tempc \@tempb \let \@tempb \@tempa \fi \ifx
  \@tempb \@empty \def\@tempb {arXiv}\fi \@ifundefined
  {mn@eprint@\@tempb}{\@tempb:\@tempc}{\expandafter \expandafter \csname
  mn@eprint@\@tempb\endcsname \expandafter{\@tempc}}}

\bibitem[\protect\citeauthoryear{{Andrae} \& {Jahnke}}{{Andrae} \&
  {Jahnke}}{2011}]{andraeyjahnke2011}
{Andrae} R.,  {Jahnke} K.,  2011, \mn@doi [\mnras]
  {10.1111/j.1365-2966.2011.19620.x}, \href
  {https://ui.adsabs.harvard.edu/abs/2011MNRAS.418.2014A} {418, 2014}

\bibitem[\protect\citeauthoryear{{Aragon-Calvo} \& {Yang}}{{Aragon-Calvo} \&
  {Yang}}{2014}]{aragon2014}
{Aragon-Calvo} M.~A.,  {Yang} L.~F.,  2014, \mn@doi [\mnras]
  {10.1093/mnrasl/slu009}, \href
  {http://adsabs.harvard.edu/abs/2014MNRAS.440L..46A} {440, L46}

\bibitem[\protect\citeauthoryear{{Arag{\'o}n-Calvo}, {Jones}, {van de Weygaert}
   \& {van der Hulst}}{{Arag{\'o}n-Calvo} et~al.}{2007a}]{aragon2007b}
{Arag{\'o}n-Calvo} M.~A.,  {Jones} B.~J.~T.,  {van de Weygaert} R.,   {van der
  Hulst} J.~M.,  2007a, \mn@doi [\aap] {10.1051/0004-6361:20077880}, \href
  {https://ui.adsabs.harvard.edu/abs/2007A&A...474..315A} {474, 315}

\bibitem[\protect\citeauthoryear{{Arag{\'o}n-Calvo}, {van de Weygaert}, {Jones}
   \& {van der Hulst}}{{Arag{\'o}n-Calvo} et~al.}{2007b}]{aragoncalvoetal2007}
{Arag{\'o}n-Calvo} M.~A.,  {van de Weygaert} R.,  {Jones} B.~J.~T.,   {van der
  Hulst} J.~M.,  2007b, \mn@doi [\apjl] {10.1086/511633}, \href
  {http://adsabs.harvard.edu/abs/2007ApJ...655L...5A} {655, L5}

\bibitem[\protect\citeauthoryear{Arag{\'o}n-Calvo, van~de Weygaert  \&
  Jones}{Arag{\'o}n-Calvo et~al.}{2010}]{aragoncalvoetal2010}
Arag{\'o}n-Calvo M.~A.,  van~de Weygaert R.,   Jones B. J.~T.,  2010, \mn@doi
  [Monthly Notices of the Royal Astronomical Society]
  {10.1111/j.1365-2966.2010.17263.x}, 408, 2163

\bibitem[\protect\citeauthoryear{{Bailin} \& {Steinmetz}}{{Bailin} \&
  {Steinmetz}}{2005}]{bailinysteinmetz2005}
{Bailin} J.,  {Steinmetz} M.,  2005, \mn@doi [\apj] {10.1086/430397}, \href
  {http://adsabs.harvard.edu/abs/2005ApJ...627..647B} {627, 647}

\bibitem[\protect\citeauthoryear{{Bardeen}, {Bond}, {Kaiser}  \&
  {Szalay}}{{Bardeen} et~al.}{1986}]{bardeenetal1986}
{Bardeen} J.~M.,  {Bond} J.~R.,  {Kaiser} N.,   {Szalay} A.~S.,  1986, \mn@doi
  [\apj] {10.1086/164143}, \href
  {https://ui.adsabs.harvard.edu/abs/1986ApJ...304...15B} {304, 15}

\bibitem[\protect\citeauthoryear{{Bett} \& {Frenk}}{{Bett} \&
  {Frenk}}{2012}]{bettyfrenk2012}
{Bett} P.~E.,  {Frenk} C.~S.,  2012, \mn@doi [\mnras]
  {10.1111/j.1365-2966.2011.20275.x}, \href
  {https://ui.adsabs.harvard.edu/abs/2012MNRAS.420.3324B} {420, 3324}

\bibitem[\protect\citeauthoryear{{Bett} \& {Frenk}}{{Bett} \&
  {Frenk}}{2016}]{bettyfrenk2016}
{Bett} P.~E.,  {Frenk} C.~S.,  2016, \mn@doi [\mnras] {10.1093/mnras/stw1395},
  \href {http://adsabs.harvard.edu/abs/2016MNRAS.461.1338B} {461, 1338}

\bibitem[\protect\citeauthoryear{{Bett}, {Eke}, {Frenk}, {Jenkins}, {Helly}  \&
  {Navarro}}{{Bett} et~al.}{2007}]{bettetal2007}
{Bett} P.,  {Eke} V.,  {Frenk} C.~S.,  {Jenkins} A.,  {Helly} J.,   {Navarro}
  J.,  2007, \mn@doi [\mnras] {10.1111/j.1365-2966.2007.11432.x}, \href
  {http://adsabs.harvard.edu/abs/2007MNRAS.376..215B} {376, 215}

\bibitem[\protect\citeauthoryear{{Blue Bird} et~al.,}{{Blue Bird}
  et~al.}{2020}]{bluebirdetal2020}
{Blue Bird} J.,  et~al., 2020, \mn@doi [\mnras] {10.1093/mnras/stz3357}, \href
  {https://ui.adsabs.harvard.edu/abs/2020MNRAS.492..153B} {492, 153}

\bibitem[\protect\citeauthoryear{{Bond}, {Kofman}  \& {Pogosyan}}{{Bond}
  et~al.}{1996}]{bondetal1996}
{Bond} J.~R.,  {Kofman} L.,   {Pogosyan} D.,  1996, \mn@doi [\nat]
  {10.1038/380603a0}, \href
  {https://ui.adsabs.harvard.edu/abs/1996Natur.380..603B} {380, 603}

\bibitem[\protect\citeauthoryear{{Borzyszkowski}, {Ludlow}  \&
  {Porciani}}{{Borzyszkowski} et~al.}{2014}]{borzyszkowskietal2014}
{Borzyszkowski} M.,  {Ludlow} A.~D.,   {Porciani} C.,  2014, \mn@doi [\mnras]
  {10.1093/mnras/stu2033}, \href
  {https://ui.adsabs.harvard.edu/abs/2014MNRAS.445.4124B} {445, 4124}

\bibitem[\protect\citeauthoryear{Cadiou, Pontzen  \& Peiris}{Cadiou
  et~al.}{2021}]{cadiouetal2021}
Cadiou C.,  Pontzen A.,   Peiris H.~V.,  2021, \mn@doi [Mon. Not. Roy. Astron.
  Soc.] {10.1093/mnras/stab440}, 502, 5480

\bibitem[\protect\citeauthoryear{{Catelan} \& {Theuns}}{{Catelan} \&
  {Theuns}}{1996a}]{catelanytheuns1996}
{Catelan} P.,  {Theuns} T.,  1996a, \mn@doi [\mnras] {10.1093/mnras/282.2.436},
  \href {https://ui.adsabs.harvard.edu/abs/1996MNRAS.282..436C} {282, 436}

\bibitem[\protect\citeauthoryear{{Catelan} \& {Theuns}}{{Catelan} \&
  {Theuns}}{1996b}]{catelanytheuns19962LPT}
{Catelan} P.,  {Theuns} T.,  1996b, \mn@doi [\mnras] {10.1093/mnras/282.2.455},
  \href {https://ui.adsabs.harvard.edu/abs/1996MNRAS.282..455C} {282, 455}

\bibitem[\protect\citeauthoryear{{Cautun}, {van de Weygaert}  \&
  {Jones}}{{Cautun} et~al.}{2013}]{cautunetal2013}
{Cautun} M.,  {van de Weygaert} R.,   {Jones} B. J.~T.,  2013, \mn@doi [\mnras]
  {10.1093/mnras/sts416}, \href
  {https://ui.adsabs.harvard.edu/abs/2013MNRAS.429.1286C} {429, 1286}

\bibitem[\protect\citeauthoryear{{Cautun}, {van de Weygaert}, {Jones}  \&
  {Frenk}}{{Cautun} et~al.}{2014}]{cautunetal2014}
{Cautun} M.,  {van de Weygaert} R.,  {Jones} B. J.~T.,   {Frenk} C.~S.,  2014,
  \mn@doi [\mnras] {10.1093/mnras/stu768}, \href
  {https://ui.adsabs.harvard.edu/abs/2014MNRAS.441.2923C} {441, 2923}

\bibitem[\protect\citeauthoryear{{Cervantes-Sodi}, {Hernandez}  \&
  {Park}}{{Cervantes-Sodi} et~al.}{2010}]{cervantessodietal2010}
{Cervantes-Sodi} B.,  {Hernandez} X.,   {Park} C.,  2010, \mn@doi [\mnras]
  {10.1111/j.1365-2966.2009.16001.x}, \href
  {https://ui.adsabs.harvard.edu/abs/2010MNRAS.402.1807C} {402, 1807}

\bibitem[\protect\citeauthoryear{{Codis}, {Pichon}, {Devriendt}, {Slyz},
  {Pogosyan}, {Dubois}  \& {Sousbie}}{{Codis} et~al.}{2012}]{codisetal2012}
{Codis} S.,  {Pichon} C.,  {Devriendt} J.,  {Slyz} A.,  {Pogosyan} D.,
  {Dubois} Y.,   {Sousbie} T.,  2012, \mn@doi [\mnras]
  {10.1111/j.1365-2966.2012.21636.x}, \href
  {https://ui.adsabs.harvard.edu/abs/2012MNRAS.427.3320C} {427, 3320}

\bibitem[\protect\citeauthoryear{{Codis}, {Pichon}  \& {Pogosyan}}{{Codis}
  et~al.}{2015}]{codisetal2015}
{Codis} S.,  {Pichon} C.,   {Pogosyan} D.,  2015, \mn@doi [\mnras]
  {10.1093/mnras/stv1570}, \href
  {http://adsabs.harvard.edu/abs/2015MNRAS.452.3369C} {452, 3369}

\bibitem[\protect\citeauthoryear{{Contreras}, {Chaves-Montero}, {Zennaro}  \&
  {Angulo}}{{Contreras} et~al.}{2021}]{contrerasetal2021}
{Contreras} S.,  {Chaves-Montero} J.,  {Zennaro} M.,   {Angulo} R.~E.,  2021,
  \mn@doi [\mnras] {10.1093/mnras/stab2367}, \href
  {https://ui.adsabs.harvard.edu/abs/2021MNRAS.507.3412C} {507, 3412}

\bibitem[\protect\citeauthoryear{{Copeland}, {Taylor}  \& {Hall}}{{Copeland}
  et~al.}{2020}]{copelandetal2020}
{Copeland} D.,  {Taylor} A.,   {Hall} A.,  2020, \mn@doi [\mnras]
  {10.1093/mnras/staa314}, \href
  {https://ui.adsabs.harvard.edu/abs/2020MNRAS.tmp..302C} {}

\bibitem[\protect\citeauthoryear{{Crittenden}, {Natarajan}, {Pen}  \&
  {Theuns}}{{Crittenden} et~al.}{2001}]{crittendenetal2001}
{Crittenden} R.~G.,  {Natarajan} P.,  {Pen} U.-L.,   {Theuns} T.,  2001,
  \mn@doi [\apj] {10.1086/322370}, \href
  {https://ui.adsabs.harvard.edu/abs/2001ApJ...559..552C} {559, 552}

\bibitem[\protect\citeauthoryear{{Dalcanton}, {Spergel}  \&
  {Summers}}{{Dalcanton} et~al.}{1997}]{dalcantonetal1997}
{Dalcanton} J.~J.,  {Spergel} D.~N.,   {Summers} F.~J.,  1997, \mn@doi [\apj]
  {10.1086/304182}, \href
  {https://ui.adsabs.harvard.edu/abs/1997ApJ...482..659D} {482, 659}

\bibitem[\protect\citeauthoryear{{Desai} \& {Ryden}}{{Desai} \&
  {Ryden}}{2022}]{desaiyryden2022}
{Desai} D.~D.,  {Ryden} B.~S.,  2022, \mn@doi [\apj]
  {10.3847/1538-4357/ac83a8}, \href
  {https://ui.adsabs.harvard.edu/abs/2022ApJ...936...25D} {936, 25}

\bibitem[\protect\citeauthoryear{{Doroshkevich}}{{Doroshkevich}}{1970}]{doroshkevich1970}
{Doroshkevich} A.~G.,  1970, Astrofizika, \href
  {http://adsabs.harvard.edu/abs/1970Afz.....6..581D} {6, 581}

\bibitem[\protect\citeauthoryear{{Dutton} et~al.,}{{Dutton}
  et~al.}{2011}]{duttonetal2011}
{Dutton} A.~A.,  et~al., 2011, \mn@doi [\mnras]
  {10.1111/j.1365-2966.2010.17555.x}, \href
  {https://ui.adsabs.harvard.edu/abs/2011MNRAS.410.1660D} {410, 1660}

\bibitem[\protect\citeauthoryear{{Ebrahimian} \& {Abolhasani}}{{Ebrahimian} \&
  {Abolhasani}}{2021}]{ebrahimianetal2021}
{Ebrahimian} E.,  {Abolhasani} A.~A.,  2021, \mn@doi [\apj]
  {10.3847/1538-4357/abd6eb}, \href
  {https://ui.adsabs.harvard.edu/abs/2021ApJ...912...57E} {912, 57}

\bibitem[\protect\citeauthoryear{{Efstathiou} \& {Jones}}{{Efstathiou} \&
  {Jones}}{1979}]{efstathiouyjones1979}
{Efstathiou} G.,  {Jones} B.~J.~T.,  1979, \mn@doi [\mnras]
  {10.1093/mnras/186.2.133}, \href
  {https://ui.adsabs.harvard.edu/abs/1979MNRAS.186..133E} {186, 133}

\bibitem[\protect\citeauthoryear{{Einasto}, {J{\~o}eveer}  \&
  {Kaasik}}{{Einasto} et~al.}{1977}]{einastoetal1977}
{Einasto} J.,  {J{\~o}eveer} M.,   {Kaasik} A.,  1977, Soviet Astronomy
  Letters, \href {https://ui.adsabs.harvard.edu/abs/1977SvAL....3..185E} {3,
  185}

\bibitem[\protect\citeauthoryear{{Fabbian}, {Lewis}  \& {Beck}}{{Fabbian}
  et~al.}{2019}]{fabbianetal2019}
{Fabbian} G.,  {Lewis} A.,   {Beck} D.,  2019, \mn@doi [\jcap]
  {10.1088/1475-7516/2019/10/057}, \href
  {https://ui.adsabs.harvard.edu/abs/2019JCAP...10..057F} {2019, 057}

\bibitem[\protect\citeauthoryear{{Fall}}{{Fall}}{1979}]{fall1979}
{Fall} S.~M.,  1979, \mn@doi [\nat] {10.1038/281200a0}, \href
  {https://ui.adsabs.harvard.edu/abs/1979Natur.281..200F} {281, 200}

\bibitem[\protect\citeauthoryear{{Feldbrugge} \& {van de
  Weygaert}}{{Feldbrugge} \& {van de Weygaert}}{2024}]{feldbruggeetal2024}
{Feldbrugge} J.,  {van de Weygaert} R.,  2024, \mn@doi [arXiv e-prints]
  {10.48550/arXiv.2405.20475}, \href
  {https://ui.adsabs.harvard.edu/abs/2024arXiv240520475F} {p. arXiv:2405.20475}

\bibitem[\protect\citeauthoryear{{Firmani} \& {Avila-Reese}}{{Firmani} \&
  {Avila-Reese}}{2009}]{firmaniyavilareese2009}
{Firmani} C.,  {Avila-Reese} V.,  2009, \mn@doi [\mnras]
  {10.1111/j.1365-2966.2009.14844.x}, \href
  {https://ui.adsabs.harvard.edu/abs/2009MNRAS.396.1675F} {396, 1675}

\bibitem[\protect\citeauthoryear{{Florack}, {ter Haar Romeny}, {Koenderink}  \&
  {Viergever}}{{Florack} et~al.}{1992}]{florack1992}
{Florack} L.,  {ter Haar Romeny} B.,  {Koenderink} J.,   {Viergever} M.,  1992,
  Image and Vision Computing, 10, 376

\bibitem[\protect\citeauthoryear{{Forero-Romero}, {Contreras}  \&
  {Padilla}}{{Forero-Romero} et~al.}{2014}]{foreroromeroetal2014}
{Forero-Romero} J.~E.,  {Contreras} S.,   {Padilla} N.,  2014, \mn@doi [\mnras]
  {10.1093/mnras/stu1150}, \href
  {http://adsabs.harvard.edu/abs/2014MNRAS.443.1090F} {443, 1090}

\bibitem[\protect\citeauthoryear{{Ganeshaiah Veena}, {Cautun}, {van de
  Weygaert}, {Tempel}, {Jones}, {Rieder}  \& {Frenk}}{{Ganeshaiah Veena}
  et~al.}{2018}]{veenaetal2018}
{Ganeshaiah Veena} P.,  {Cautun} M.,  {van de Weygaert} R.,  {Tempel} E.,
  {Jones} B. J.~T.,  {Rieder} S.,   {Frenk} C.~S.,  2018, \mn@doi [\mnras]
  {10.1093/mnras/sty2270}, \href
  {https://ui.adsabs.harvard.edu/abs/2018MNRAS.481..414G} {481, 414}

\bibitem[\protect\citeauthoryear{Ganeshaiah~Veena, Cautun, van~de Weygaert,
  Tempel  \& Frenk}{Ganeshaiah~Veena et~al.}{2021a}]{veenaetal2019}
Ganeshaiah~Veena P.,  Cautun M.,  van~de Weygaert R.,  Tempel E.,   Frenk
  C.~S.,  2021a, \mn@doi [MNRAS] {10.1093/mnras/stab411}

\bibitem[\protect\citeauthoryear{{Ganeshaiah Veena}, {Cautun}, {van de
  Weygaert}, {Tempel}  \& {Frenk}}{{Ganeshaiah Veena}
  et~al.}{2021b}]{veenaetal2021}
{Ganeshaiah Veena} P.,  {Cautun} M.,  {van de Weygaert} R.,  {Tempel} E.,
  {Frenk} C.~S.,  2021b, \mn@doi [\mnras] {10.1093/mnras/stab411}, \href
  {https://ui.adsabs.harvard.edu/abs/2021MNRAS.503.2280G} {503, 2280}

\bibitem[\protect\citeauthoryear{{Gao} \& {White}}{{Gao} \&
  {White}}{2007}]{gaoywhite2007}
{Gao} L.,  {White} S.~D.~M.,  2007, \mn@doi [\mnras]
  {10.1111/j.1745-3933.2007.00292.x}, \href
  {http://adsabs.harvard.edu/abs/2007MNRAS.377L...5G} {377, L5}

\bibitem[\protect\citeauthoryear{{Gao}, {Springel}  \& {White}}{{Gao}
  et~al.}{2005}]{gaoetal2005}
{Gao} L.,  {Springel} V.,   {White} S. D.~M.,  2005, \mn@doi [\mnras]
  {10.1111/j.1745-3933.2005.00084.x}, \href
  {https://ui.adsabs.harvard.edu/abs/2005MNRAS.363L..66G} {363, L66}

\bibitem[\protect\citeauthoryear{{Hahn}, {Carollo}, {Porciani}  \&
  {Dekel}}{{Hahn} et~al.}{2007}]{hahnetal2007}
{Hahn} O.,  {Carollo} C.~M.,  {Porciani} C.,   {Dekel} A.,  2007, \mn@doi
  [\mnras] {10.1111/j.1365-2966.2007.12249.x}, \href
  {http://adsabs.harvard.edu/abs/2007MNRAS.381...41H} {381, 41}

\bibitem[\protect\citeauthoryear{{Hahn}, {Porciani}, {Dekel}  \&
  {Carollo}}{{Hahn} et~al.}{2009}]{hahnetal2009}
{Hahn} O.,  {Porciani} C.,  {Dekel} A.,   {Carollo} C.~M.,  2009, \mn@doi
  [\mnras] {10.1111/j.1365-2966.2009.15271.x}, \href
  {https://ui.adsabs.harvard.edu/abs/2009MNRAS.398.1742H} {398, 1742}

\bibitem[\protect\citeauthoryear{{Hahn}, {Teyssier}  \& {Carollo}}{{Hahn}
  et~al.}{2010}]{hahnetal2010}
{Hahn} O.,  {Teyssier} R.,   {Carollo} C.~M.,  2010, \mn@doi [\mnras]
  {10.1111/j.1365-2966.2010.16494.x}, \href
  {https://ui.adsabs.harvard.edu/abs/2010MNRAS.405..274H} {405, 274}

\bibitem[\protect\citeauthoryear{{Heavens} \& {Peacock}}{{Heavens} \&
  {Peacock}}{1988}]{heavensypeacock1988}
{Heavens} A.,  {Peacock} J.,  1988, \mn@doi [\mnras] {10.1093/mnras/232.2.339},
  \href {http://adsabs.harvard.edu/abs/1988MNRAS.232..339H} {232, 339}

\bibitem[\protect\citeauthoryear{{Hikage} et~al.,}{{Hikage}
  et~al.}{2019}]{hikageetal2019}
{Hikage} C.,  et~al., 2019, \mn@doi [\pasj] {10.1093/pasj/psz010}, \href
  {https://ui.adsabs.harvard.edu/abs/2019PASJ...71...43H} {71, 43}

\bibitem[\protect\citeauthoryear{{Hoyle}, {Burgers}  \& {van de Hulst}}{{Hoyle}
  et~al.}{1949}]{hoyle1949}
{Hoyle} F.,  {Burgers} J.~M.,   {van de Hulst} H.~C.,  1949, {eds., in Problems
  of Cosmical Aerodynamics, Central Air Documents Office, Dayton, p. 195}.
coso

\bibitem[\protect\citeauthoryear{{Jones}, {van de Weygaert}  \&
  {Arag{\'o}n-Calvo}}{{Jones} et~al.}{2010}]{jonesetal2010}
{Jones} B. J.~T.,  {van de Weygaert} R.,   {Arag{\'o}n-Calvo} M.~A.,  2010,
  \mn@doi [\mnras] {10.1111/j.1365-2966.2010.17202.x}, \href
  {https://ui.adsabs.harvard.edu/abs/2010MNRAS.408..897J} {408, 897}

\bibitem[\protect\citeauthoryear{{Karachentsev} \& {Zozulia}}{{Karachentsev} \&
  {Zozulia}}{2023}]{karachentsevetal2023}
{Karachentsev} I.~D.,  {Zozulia} V.~D.,  2023, \mn@doi [\mnras]
  {10.1093/mnras/stad1279}, \href
  {https://ui.adsabs.harvard.edu/abs/2023MNRAS.522.4740K} {522, 4740}

\bibitem[\protect\citeauthoryear{Kraljic, Duckworth, Tojeiro, Alam, Bizyaev,
  Weijmans, Boardman  \& Lane}{Kraljic et~al.}{2021}]{kraljicetal2021}
Kraljic K.,  Duckworth C.,  Tojeiro R.,  Alam S.,  Bizyaev D.,  Weijmans A.-M.,
   Boardman N.~F.,   Lane R.~R.,  2021, \mn@doi [Monthly Notices of the Royal
  Astronomical Society] {10.1093/mnras/stab1109}, 504, 4626

\bibitem[\protect\citeauthoryear{{Krolewski}, {Ho}, {Chen}, {Chan}, {Tenneti},
  {Bizyaev}  \& {Kraljic}}{{Krolewski} et~al.}{2019}]{krolewskietal2019}
{Krolewski} A.,  {Ho} S.,  {Chen} Y.-C.,  {Chan} P.~F.,  {Tenneti} A.,
  {Bizyaev} D.,   {Kraljic} K.,  2019, \mn@doi [\apj]
  {10.3847/1538-4357/ab1010}, \href
  {https://ui.adsabs.harvard.edu/abs/2019ApJ...876...52K} {876, 52}

\bibitem[\protect\citeauthoryear{{Kugel} \& {van de Weygaert}}{{Kugel} \& {van
  de Weygaert}}{2024}]{kugel2024}
{Kugel} R.,  {van de Weygaert} R.,  2024, \mn@doi [arXiv e-prints]
  {10.48550/arXiv.2407.16489}, \href
  {https://ui.adsabs.harvard.edu/abs/2024arXiv240716489K} {p. arXiv:2407.16489}

\bibitem[\protect\citeauthoryear{{Laigle} et~al.,}{{Laigle}
  et~al.}{2015}]{laigleetal2015}
{Laigle} C.,  et~al., 2015, \mn@doi [\mnras] {10.1093/mnras/stu2289}, \href
  {http://adsabs.harvard.edu/abs/2015MNRAS.446.2744L} {446, 2744}

\bibitem[\protect\citeauthoryear{{Lazeyras}, {Villaescusa-Navarro}  \&
  {Viel}}{{Lazeyras} et~al.}{2021}]{lazeyrasetal2021}
{Lazeyras} T.,  {Villaescusa-Navarro} F.,   {Viel} M.,  2021, \mn@doi [\jcap]
  {10.1088/1475-7516/2021/03/022}, \href
  {https://ui.adsabs.harvard.edu/abs/2021JCAP...03..022L} {2021, 022}

\bibitem[\protect\citeauthoryear{{Lee}}{{Lee}}{2011}]{lee2011}
{Lee} J.,  2011, \mn@doi [\apj] {10.1088/0004-637X/732/2/99}, \href
  {https://ui.adsabs.harvard.edu/abs/2011ApJ...732...99L} {732, 99}

\bibitem[\protect\citeauthoryear{{Lee} \& {Erdogdu}}{{Lee} \&
  {Erdogdu}}{2007}]{leeyerdogdu2007}
{Lee} J.,  {Erdogdu} P.,  2007, \mn@doi [\apj] {10.1086/523351}, \href
  {http://adsabs.harvard.edu/abs/2007ApJ...671.1248L} {671, 1248}

\bibitem[\protect\citeauthoryear{{Lee} \& {Moon}}{{Lee} \&
  {Moon}}{2022}]{leeymoon2022}
{Lee} J.,  {Moon} J.-S.,  2022, \mn@doi [\apj] {10.3847/1538-4357/ac879d},
  \href {https://ui.adsabs.harvard.edu/abs/2022ApJ...936..119L} {936, 119}

\bibitem[\protect\citeauthoryear{{Lee} \& {Moon}}{{Lee} \&
  {Moon}}{2023}]{leeetal2023}
{Lee} J.,  {Moon} J.-S.,  2023, \mn@doi [\apjl] {10.3847/2041-8213/acdd75},
  \href {https://ui.adsabs.harvard.edu/abs/2023ApJ...951L..26L} {951, L26}

\bibitem[\protect\citeauthoryear{{Lee} \& {Pen}}{{Lee} \&
  {Pen}}{2000}]{leeypen2000}
{Lee} J.,  {Pen} U.-L.,  2000, \mn@doi [\apjl] {10.1086/312556}, \href
  {https://ui.adsabs.harvard.edu/abs/2000ApJ...532L...5L} {532, L5}

\bibitem[\protect\citeauthoryear{{Lee} \& {Pen}}{{Lee} \&
  {Pen}}{2001}]{leeypen2001}
{Lee} J.,  {Pen} U.-L.,  2001, \mn@doi [\apj] {10.1086/321472}, \href
  {https://ui.adsabs.harvard.edu/abs/2001ApJ...555..106L} {555, 106}

\bibitem[\protect\citeauthoryear{{Lee}, {Hahn}  \& {Porciani}}{{Lee}
  et~al.}{2009}]{leeetal2009}
{Lee} J.,  {Hahn} O.,   {Porciani} C.,  2009, \mn@doi [\apj]
  {10.1088/0004-637X/707/1/761}, \href
  {https://ui.adsabs.harvard.edu/abs/2009ApJ...707..761L} {707, 761}

\bibitem[\protect\citeauthoryear{Lee, Libeskind  \& Ryu}{Lee
  et~al.}{2020}]{leeetal2020}
Lee J.,  Libeskind N.~I.,   Ryu S.,  2020, \mn@doi [The Astrophysical Journal]
  {10.3847/2041-8213/aba2ee}, 898, L27

\bibitem[\protect\citeauthoryear{{Libeskind}, {Hoffman}, {Steinmetz},
  {Gottl{\"o}ber}, {Knebe}  \& {Hess}}{{Libeskind}
  et~al.}{2013}]{libeskindetal2013}
{Libeskind} N.~I.,  {Hoffman} Y.,  {Steinmetz} M.,  {Gottl{\"o}ber} S.,
  {Knebe} A.,   {Hess} S.,  2013, \mn@doi [\apj] {10.1088/2041-8205/766/2/L15},
  \href {https://ui.adsabs.harvard.edu/abs/2013ApJ...766L..15L} {766, L15}

\bibitem[\protect\citeauthoryear{{Libeskind} et~al.,}{{Libeskind}
  et~al.}{2018}]{libeskind2018}
{Libeskind} N.~I.,  et~al., 2018, \mn@doi [\mnras] {10.1093/mnras/stx1976},
  \href {http://adsabs.harvard.edu/abs/2018MNRAS.473.1195L} {473, 1195}

\bibitem[\protect\citeauthoryear{{Lindeberg}}{{Lindeberg}}{1994}]{lindeberg1994}
{Lindeberg} T.,  1994, \mn@doi [Journal of Applied Statistics]
  {10.1080/757582976}, \href
  {https://ui.adsabs.harvard.edu/abs/1994JApSt..21..225L} {21, 225}

\bibitem[\protect\citeauthoryear{{L{\'o}pez}, {Merch{\'a}n}  \&
  {Paz}}{{L{\'o}pez} et~al.}{2019}]{lopezetal2019}
{L{\'o}pez} P.,  {Merch{\'a}n} M.~E.,   {Paz} D.~J.,  2019, \mn@doi [\mnras]
  {10.1093/mnras/stz762}, \href
  {https://ui.adsabs.harvard.edu/abs/2019MNRAS.485.5244L} {485, 5244}

\bibitem[\protect\citeauthoryear{{L{\'o}pez}, {Cautun}, {Paz}, {Merch{\'a}n}
  \& {van de Weygaert}}{{L{\'o}pez} et~al.}{2021}]{lopezetal2021}
{L{\'o}pez} P.,  {Cautun} M.,  {Paz} D.,  {Merch{\'a}n} M.,   {van de Weygaert}
  R.,  2021, \mn@doi [\mnras] {10.1093/mnras/stab451}, \href
  {https://ui.adsabs.harvard.edu/abs/2021MNRAS.502.5528L} {502, 5528}

\bibitem[\protect\citeauthoryear{{Ludlow}, {Borzyszkowski}  \&
  {Porciani}}{{Ludlow} et~al.}{2014}]{Ludlow2014}
{Ludlow} A.~D.,  {Borzyszkowski} M.,   {Porciani} C.,  2014, \mn@doi [\mnras]
  {10.1093/mnras/stu2021}, \href
  {https://ui.adsabs.harvard.edu/abs/2014MNRAS.445.4110L} {445, 4110}

\bibitem[\protect\citeauthoryear{{López} \& {Merchán}}{{López} \&
  {Merchán}}{2023}]{lopez2023}
{López} P.,  {Merchán} M.~E.,  2023, Momento angular y dinámica interna de
  halos de materia.
s.n., S.l., \url {http://hdl.handle.net/11086/551207}

\bibitem[\protect\citeauthoryear{{McEwen} \& {Weinberg}}{{McEwen} \&
  {Weinberg}}{2018}]{mcewenetal2018}
{McEwen} J.~E.,  {Weinberg} D.~H.,  2018, \mn@doi [\mnras]
  {10.1093/mnras/sty882}, \href
  {https://ui.adsabs.harvard.edu/abs/2018MNRAS.477.4348M} {477, 4348}

\bibitem[\protect\citeauthoryear{{Mo}, {Mao}  \& {White}}{{Mo}
  et~al.}{1998}]{momaoywhite1998}
{Mo} H.~J.,  {Mao} S.,   {White} S. D.~M.,  1998, \mn@doi [\mnras]
  {10.1046/j.1365-8711.1998.01227.x}, \href
  {https://ui.adsabs.harvard.edu/abs/1998MNRAS.295..319M} {295, 319}

\bibitem[\protect\citeauthoryear{{Montero-Dorta}, {Contreras}, {Artale},
  {Rodriguez}  \& {Favole}}{{Montero-Dorta}
  et~al.}{2024}]{monterodortaetal2024}
{Montero-Dorta} A.~D.,  {Contreras} S.,  {Artale} M.~C.,  {Rodriguez} F.,
  {Favole} G.,  2024, \mn@doi [arXiv e-prints] {10.48550/arXiv.2410.18319},
  \href {https://ui.adsabs.harvard.edu/abs/2024arXiv241018319M} {p.
  arXiv:2410.18319}

\bibitem[\protect\citeauthoryear{{Moon} \& {Lee}}{{Moon} \&
  {Lee}}{2023}]{moonylee2023}
{Moon} J.-S.,  {Lee} J.,  2023, \mn@doi [\apj] {10.3847/1538-4357/acd7ed},
  \href {https://ui.adsabs.harvard.edu/abs/2023ApJ...952...82M} {952, 82}

\bibitem[\protect\citeauthoryear{Moon \& Lee}{Moon \& Lee}{2024}]{moonylee2024}
Moon J.-S.,  Lee J.,  2024, Galaxy Spin Transition Driven by the Misalignments
  between the Protogalaxy Inertia and Initial Tidal Tensors (\mn@eprint {arXiv}
  {2401.11707}), \url {https://arxiv.org/abs/2401.11707}

\bibitem[\protect\citeauthoryear{{Navarro}, {Abadi}  \& {Steinmetz}}{{Navarro}
  et~al.}{2004}]{navarroetal2004}
{Navarro} J.~F.,  {Abadi} M.~G.,   {Steinmetz} M.,  2004, \mn@doi [\apjl]
  {10.1086/424902}, \href
  {https://ui.adsabs.harvard.edu/abs/2004ApJ...613L..41N} {613, L41}

\bibitem[\protect\citeauthoryear{{Neyrinck}, {Aragon-Calvo}, {Falck}, {Szalay}
  \& {Wang}}{{Neyrinck} et~al.}{2020}]{neyrincketal2019}
{Neyrinck} M.,  {Aragon-Calvo} M.~A.,  {Falck} B.,  {Szalay} A.~S.,   {Wang}
  J.,  2020, \mn@doi [The Open Journal of Astrophysics]
  {10.21105/astro.1904.03201}, \href
  {https://ui.adsabs.harvard.edu/abs/2020OJAp....3E...3N} {3, 3}

\bibitem[\protect\citeauthoryear{{Neyrinck}, {Arag{\'o}n-Calvo}  \&
  {Szapudi}}{{Neyrinck} et~al.}{2025}]{neyrincketal2025}
{Neyrinck} M.,  {Arag{\'o}n-Calvo} M.,   {Szapudi} I.,  2025, \mn@doi [arXiv
  e-prints] {10.48550/arXiv.2503.21015}, \href
  {https://ui.adsabs.harvard.edu/abs/2025arXiv250321015N} {p. arXiv:2503.21015}

\bibitem[\protect\citeauthoryear{{Paz}, {Lambas}, {Padilla}  \&
  {Merch{\'a}n}}{{Paz} et~al.}{2006}]{pazetal2006}
{Paz} D.~J.,  {Lambas} D.~G.,  {Padilla} N.,   {Merch{\'a}n} M.,  2006, \mn@doi
  [\mnras] {10.1111/j.1365-2966.2005.09934.x}, \href
  {http://adsabs.harvard.edu/abs/2006MNRAS.366.1503P} {366, 1503}

\bibitem[\protect\citeauthoryear{{Paz}, {Stasyszyn}  \& {Padilla}}{{Paz}
  et~al.}{2008}]{pazetal2008}
{Paz} D.~J.,  {Stasyszyn} F.,   {Padilla} N.~D.,  2008, \mn@doi [\mnras]
  {10.1111/j.1365-2966.2008.13655.x}, \href
  {http://adsabs.harvard.edu/abs/2008MNRAS.389.1127P} {389, 1127}

\bibitem[\protect\citeauthoryear{{Peebles}}{{Peebles}}{1969}]{peebles1969}
{Peebles} P.~J.~E.,  1969, \mn@doi [\apj] {10.1086/149876}, \href
  {http://adsabs.harvard.edu/abs/1969ApJ...155..393P} {155, 393}

\bibitem[\protect\citeauthoryear{{Peebles}}{{Peebles}}{1971}]{peebles1971}
{Peebles} P.~J.~E.,  1971, \aap, \href
  {https://ui.adsabs.harvard.edu/abs/1971A&A....11..377P} {11, 377}

\bibitem[\protect\citeauthoryear{{Pereyra}, {Sgr{\'o}}, {Merch{\'a}n},
  {Stasyszyn}  \& {Paz}}{{Pereyra} et~al.}{2020}]{pereyraetal2020}
{Pereyra} L.~A.,  {Sgr{\'o}} M.~A.,  {Merch{\'a}n} M.~E.,  {Stasyszyn} F.~A.,
  {Paz} D.~J.,  2020, \mn@doi [\mnras] {10.1093/mnras/staa3112}, \href
  {https://ui.adsabs.harvard.edu/abs/2020MNRAS.499.4876P} {499, 4876}

\bibitem[\protect\citeauthoryear{{Planck Collaboration} et~al.,}{{Planck
  Collaboration} et~al.}{2020}]{plankcollaboration2018}
{Planck Collaboration} et~al., 2020, \mn@doi [A\&A]
  {10.1051/0004-6361/201833910}, 641, A6

\bibitem[\protect\citeauthoryear{{Porciani}, {Dekel}  \& {Hoffman}}{{Porciani}
  et~al.}{2002a}]{porcianietal2002a}
{Porciani} C.,  {Dekel} A.,   {Hoffman} Y.,  2002a, \mn@doi [\mnras]
  {10.1046/j.1365-8711.2002.05305.x}, \href
  {http://adsabs.harvard.edu/abs/2002MNRAS.332..325P} {332, 325}

\bibitem[\protect\citeauthoryear{{Porciani}, {Dekel}  \& {Hoffman}}{{Porciani}
  et~al.}{2002b}]{porcianietal2002b}
{Porciani} C.,  {Dekel} A.,   {Hoffman} Y.,  2002b, \mn@doi [\mnras]
  {10.1046/j.1365-8711.2002.05306.x}, \href
  {http://adsabs.harvard.edu/abs/2002MNRAS.332..339P} {332, 339}

\bibitem[\protect\citeauthoryear{{Rodriguez}, {Merch{\'a}n}  \&
  {Sgr{\'o}}}{{Rodriguez} et~al.}{2015}]{rodriguezetal2015}
{Rodriguez} F.,  {Merch{\'a}n} M.,   {Sgr{\'o}} M.~A.,  2015, \mn@doi [\aap]
  {10.1051/0004-6361/201525798}, \href
  {https://ui.adsabs.harvard.edu/abs/2015A&A...580A..86R} {580, A86}

\bibitem[\protect\citeauthoryear{{Rong} \& {Wang}}{{Rong} \&
  {Wang}}{2025}]{rongetal2025b}
{Rong} Y.,  {Wang} P.,  2025, \mn@doi [\apj] {10.3847/1538-4357/adc4de}, \href
  {https://ui.adsabs.harvard.edu/abs/2025ApJ...983..122R} {983, 122}

\bibitem[\protect\citeauthoryear{Schaefer}{Schaefer}{2009}]{schafer2009}
Schaefer B.~M.,  2009, \mn@doi [Int. J. Mod. Phys. D]
  {10.1142/S0218271809014388}, 18, 173

\bibitem[\protect\citeauthoryear{{Sch{\"a}fer} \& {Merkel}}{{Sch{\"a}fer} \&
  {Merkel}}{2012}]{schaferetal2012}
{Sch{\"a}fer} B.~M.,  {Merkel} P.~M.,  2012, \mn@doi [\mnras]
  {10.1111/j.1365-2966.2011.20224.x}, \href
  {https://ui.adsabs.harvard.edu/abs/2012MNRAS.421.2751S} {421, 2751}

\bibitem[\protect\citeauthoryear{{Shandarin} \& {Zeldovich}}{{Shandarin} \&
  {Zeldovich}}{1989}]{shandarinyzedovich1989}
{Shandarin} S.~F.,  {Zeldovich} Y.~B.,  1989, \mn@doi [Reviews of Modern
  Physics] {10.1103/RevModPhys.61.185}, \href
  {https://ui.adsabs.harvard.edu/abs/1989RvMP...61..185S} {61, 185}

\bibitem[\protect\citeauthoryear{Sheth \& Tormen}{Sheth \&
  Tormen}{2002}]{shethytormen2002}
Sheth R.~K.,  Tormen G.,  2002, \mn@doi [Monthly Notices of the Royal
  Astronomical Society] {10.1046/j.1365-8711.2002.04950.x}, 329, 61

\bibitem[\protect\citeauthoryear{{Sheth} \& {Tormen}}{{Sheth} \&
  {Tormen}}{2004}]{shethytormen2004}
{Sheth} R.~K.,  {Tormen} G.,  2004, \mn@doi [\mnras]
  {10.1111/j.1365-2966.2004.07733.x}, \href
  {https://ui.adsabs.harvard.edu/abs/2004MNRAS.350.1385S} {350, 1385}

\bibitem[\protect\citeauthoryear{{Somerville} et~al.,}{{Somerville}
  et~al.}{2008}]{somervilleetal2008}
{Somerville} R.~S.,  et~al., 2008, \mn@doi [\apj] {10.1086/523661}, \href
  {https://ui.adsabs.harvard.edu/abs/2008ApJ...672..776S} {672, 776}

\bibitem[\protect\citeauthoryear{{Sousbie}, {Pichon}, {Colombi}, {Novikov}  \&
  {Pogosyan}}{{Sousbie} et~al.}{2008}]{sousbieetal2008}
{Sousbie} T.,  {Pichon} C.,  {Colombi} S.,  {Novikov} D.,   {Pogosyan} D.,
  2008, \mn@doi [\mnras] {10.1111/j.1365-2966.2007.12685.x}, \href
  {https://ui.adsabs.harvard.edu/abs/2008MNRAS.383.1655S} {383, 1655}

\bibitem[\protect\citeauthoryear{{Springel}}{{Springel}}{2005}]{springelgadget2005}
{Springel} V.,  2005, \mn@doi [\mnras] {10.1111/j.1365-2966.2005.09655.x},
  \href {http://adsabs.harvard.edu/abs/2005MNRAS.364.1105S} {364, 1105}

\bibitem[\protect\citeauthoryear{{Sugerman}, {Summers}  \&
  {Kamionkowski}}{{Sugerman} et~al.}{2000}]{sugermanetal2000}
{Sugerman} B.,  {Summers} F.~J.,   {Kamionkowski} M.,  2000, \mn@doi [\mnras]
  {10.1046/j.1365-8711.2000.03107.x}, \href
  {http://adsabs.harvard.edu/abs/2000MNRAS.311..762S} {311, 762}

\bibitem[\protect\citeauthoryear{{Tempel} \& {Libeskind}}{{Tempel} \&
  {Libeskind}}{2013}]{tempelylibeskind2013}
{Tempel} E.,  {Libeskind} N.~I.,  2013, \mn@doi [\apjl]
  {10.1088/2041-8205/775/2/L42}, \href
  {http://adsabs.harvard.edu/abs/2013ApJ...775L..42T} {775, L42}

\bibitem[\protect\citeauthoryear{{Troxel} \& {Ishak}}{{Troxel} \&
  {Ishak}}{2015}]{troxelyishak2014}
{Troxel} M.~A.,  {Ishak} M.,  2015, \mn@doi [\physrep]
  {10.1016/j.physrep.2014.11.001}, \href
  {https://ui.adsabs.harvard.edu/abs/2015PhR...558....1T} {558, 1}

\bibitem[\protect\citeauthoryear{{Vitvitska}, {Klypin}, {Kravtsov}, {Wechsler},
  {Primack}  \& {Bullock}}{{Vitvitska} et~al.}{2002}]{vitvitskaetal2002}
{Vitvitska} M.,  {Klypin} A.~A.,  {Kravtsov} A.~V.,  {Wechsler} R.~H.,
  {Primack} J.~R.,   {Bullock} J.~S.,  2002, \mn@doi [\apj] {10.1086/344361},
  \href {http://adsabs.harvard.edu/abs/2002ApJ...581..799V} {581, 799}

\bibitem[\protect\citeauthoryear{{Wang} \& {Kang}}{{Wang} \&
  {Kang}}{2017}]{wangykang2017}
{Wang} P.,  {Kang} X.,  2017, \mn@doi [\mnras] {10.1093/mnrasl/slx038}, \href
  {https://ui.adsabs.harvard.edu/abs/2017MNRAS.468L.123W} {468, L123}

\bibitem[\protect\citeauthoryear{{Wechsler}, {Zentner}, {Bullock}, {Kravtsov}
  \& {Allgood}}{{Wechsler} et~al.}{2006}]{wechsleretal2006}
{Wechsler} R.~H.,  {Zentner} A.~R.,  {Bullock} J.~S.,  {Kravtsov} A.~V.,
  {Allgood} B.,  2006, \mn@doi [\apj] {10.1086/507120}, \href
  {https://ui.adsabs.harvard.edu/abs/2006ApJ...652...71W} {652, 71}

\bibitem[\protect\citeauthoryear{{Welker} et~al.,}{{Welker}
  et~al.}{2020}]{welkeretal2019}
{Welker} C.,  et~al., 2020, \mn@doi [\mnras] {10.1093/mnras/stz2860}, \href
  {https://ui.adsabs.harvard.edu/abs/2020MNRAS.491.2864W} {491, 2864}

\bibitem[\protect\citeauthoryear{{White}}{{White}}{1984}]{white1984}
{White} S.~D.~M.,  1984, \mn@doi [\apj] {10.1086/162573}, \href
  {http://adsabs.harvard.edu/abs/1984ApJ...286...38W} {286, 38}

\bibitem[\protect\citeauthoryear{{Zel'dovich}}{{Zel'dovich}}{1970}]{zeldovich1970}
{Zel'dovich} Y.~B.,  1970, \aap, \href
  {http://adsabs.harvard.edu/abs/1970A%26A.....5...84Z} {5, 84}

\bibitem[\protect\citeauthoryear{{Zentner}, {Hearin}  \& {van den
  Bosch}}{{Zentner} et~al.}{2014}]{zentneretal2014}
{Zentner} A.~R.,  {Hearin} A.~P.,   {van den Bosch} F.~C.,  2014, \mn@doi
  [\mnras] {10.1093/mnras/stu1383}, \href
  {https://ui.adsabs.harvard.edu/abs/2014MNRAS.443.3044Z} {443, 3044}

\bibitem[\protect\citeauthoryear{{Zhang}, {Yang}, {Wang}, {Wang}, {Luo}, {Mo}
  \& {van den Bosch}}{{Zhang} et~al.}{2015}]{zhangetal2015}
{Zhang} Y.,  {Yang} X.,  {Wang} H.,  {Wang} L.,  {Luo} W.,  {Mo} H.~J.,   {van
  den Bosch} F.~C.,  2015, \mn@doi [\apj] {10.1088/0004-637X/798/1/17}, \href
  {http://adsabs.harvard.edu/abs/2015ApJ...798...17Z} {798, 17}

\bibitem[\protect\citeauthoryear{{van de Weygaert} \& {Bond}}{{van de Weygaert}
  \& {Bond}}{2008}]{weygaert&bond2008}
{van de Weygaert} R.,  {Bond} J.~R.,  2008, {Observations and Morphology of the
  Cosmic Web}.
coso, p.~24, \mn@doi{10.1007/978-1-4020-6941-3_11}

\makeatother
\end{thebibliography}




\appendix

\section{Cosmic web identification with \nexus}
\label{app:nexus}


\begin{figure*}
	\includegraphics[width=1.8\columnwidth]{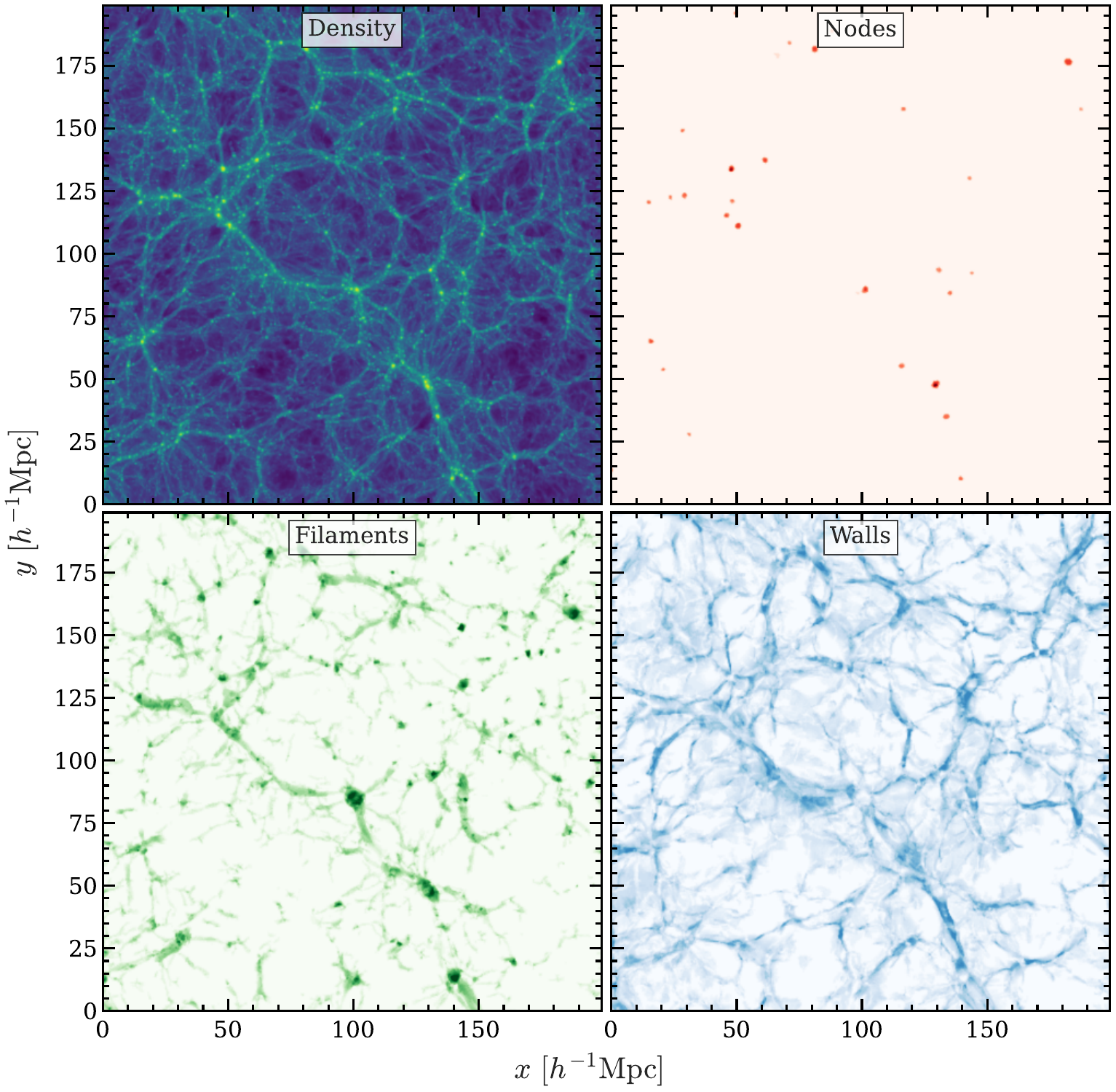}
   \caption{Example of structure identification with the \nexus{} algorithm in a $200^{-1}\mathrm{Mpc}$ side and $10h^{-1}\mathrm{Mpc}$ thick slice of the simulation at redshift $z=0$. The upper left panel shows the matter density field $\delta(\mathbf{x}) + 1$, where brighter regions correspond to higher densities. The remaining panels display the \nexus{} signature $\delta^{\text{NEXUS}}_j(\mathbf{x})$, averaged over the slice thickness, for the three cosmic web components: nodes (upper right), filaments (bottom left) and walls (bottom right). In these panels, darker colors indicate values closer to unity.}
   \label{fig:nexus}
 \end{figure*}

The \nexus{} suite of cosmic web identifiers represents an elaboration and extension of the original Multiscale Morphology Filter \citep{aragoncalvoetal2007,aragon2007b,aragoncalvoetal2010} algorithm and was developed with the goal of obtaining a more physically motivated and robust method (for short summaries see also \cite{libeskind2018,kugel2024}). \nexus{} is the principal representative of the full \nexus{} suite of cosmic web identifiers \citep[see][]{cautunetal2013}. These include the options for multiscale morphology identifiers on the basis of the linear density, the logarithmic density, the velocity divergence, the velocity shear and tidal force field. \nexus{} has incorporated these options in a versatile code for the analysis of cosmic web structure and dynamics following the realization that they are significant physical influences in shaping the cosmic mass distribution into the complexity of the cosmic web.

\subsection{Hessian Geometry and Morphological Identity}
The basic setup of \MMF/\nexus{} is that of defining a four-dimensional scale-space representation of the input field $f(\vec{x})$. In nearly all implementations this achieved by means of a Gaussian filtering of $f(\vec{x})$ over a set of
scales $[R_0,R_1,...,R_N]$, 
\begin{equation}
    f_{R_n}(\vec{x}) = \int \frac{{\rm d}^3k}{(2\pi)^3} e^{-k^2R_n^2/2} \hat{f}(\vec{k})  e^{i\vec{k}\cdot\vec{x}} ,
    \label{eq:filtered_field}
\end{equation}
\noindent where $\hat{f}(\vec{k})$ is the Fourier transform of the input field $f(\vec{x})$. The Hessian $H_{ij,R_n}(\vec{x})$ of the filtered field on the scale $R_n$ is computed in Fourier space on the basis of the corresponding Fourier components $\hat{H}_{ij,R_n}(\vec{k})$,
\begin{eqnarray}
    H_{ij,R_n}(\vec{x})&\,=\,&R_n^2 \; \frac{\partial^2f_{R_n}(\vec{x})}{\partial x_i\partial x_j}\,.
    \ \\
    \hat{H}_{ij,R_n}(\vec{k})&\,=\,& -k_ik_j R_n^2 \hat{f}(\vec{k}) e^{-k^2R_n^2/2}\,.\nonumber
    \label{eq:hessian_general}
\end{eqnarray}
\noindent Note that the definition for the Hessian includes the normalization term $R_n^2$. The key element of the \MMF/\nexus{} formalism is the morphological information contained in the eigenvalues of the Hessian matrix, $h_1 \le h_2 \le h_3$. By applying a set of morphology filters on these scaled eigenvalues \citep[see][]{aragon2007b,cautunetal2013} this is translated into a scale dependent environment signature $\mathcal{S}_{R_n}(\mathbf{ x})$ that represents the geometry at the corresponding scale. 

\subsection{Scale Space and Multiscale Structure}
To analyse the multiscale nature of the cosmic web, the \textit{Scale-Space} representation of the cosmic mass distributions produces a sequence of copies of the data having different resolutions \citep{florack1992,lindeberg1994}. At each location $\vec{x}$ in the probed volume, it involves an extra dimension, scale, that yields the eigenvalues of the Hessian filtered at the corresponding (Gaussian) scale and the scale-dependent environment signature $\mathcal{S}_{R_n}(\mathbf{ x})$.

A feature searching algorithm is applied to the combined set of scaled copies in order to identify the scale at which, locally, we find the strongest morphological signature. It involves the combination of the complete set of scale-dependent environmental signatures to find the maximum signature for all scales:
\begin{equation}
    \mathcal{S}(\mathbf{ x}) = \max\limits_{\text{levels n}} \mathcal{S}_{R_n}(\mathbf{ x}).
    \label{eq:NEXUSsig}
\end{equation}

\subsection{Signature \& Versions} 
The final step in the \MMF/\nexus{} procedure involves the use of criteria to find the threshold signature that discriminates between valid and invalid morphological detections. Signature values larger than the threshold correspond to real structures while the rest are spurious detections. Different implementations and versions of the \MMF/\nexus{} technique may differ in the definition of the threshold values. 

The final outcome of the \MMF/\nexus{} procedure is a field which at each location $\vec{x}$ specifies what the local morphological signature is, cluster node, filaments, wall or void. The resulting field $\delta^{\text{NEXUS}}_j(\mathbf{ x})$ is zero when the volume is not identified as cosmic web element $j$ and is one when the volume element is identified as element $j$. Here $j$ is either filaments, nodes or walls. In this identification we also intrinsically include the identification for voids which is defined as the volume elements that are neither a filament, node or wall.

The principal technique, and corresponding philosophy, has branched in several elaborations and developments of the original Multiscale Morphology Filter \citep{cautunetal2014,aragon2014}. The \nexus{} suite of cosmic web identifiers \citep{cautunetal2014} has extended the \MMF{} formalism to a substantially wider range of physical agents involved in the formation of the cosmic web, along with a substantially firmer foundation for the criteria used in identifying the various web-like structures.  These include the options for multiscale morphology identifiers on the basis of the linear density, the logarithmic density, the velocity divergence, the velocity shear and tidal force field. \nexus{} has incorporated these options in a versatile code for the analysis of cosmic web structure and dynamics following the realisation that they are significant physical influences in shaping the cosmic mass distribution into the complexity of the cosmic web.

\subsection{NEXUS+}
\nexus{} is the principal representative of the full \nexus{} suite of cosmic web identifiers \citep[see][]{cautunetal2013}. It is the density field \nexus{} version with the highest dynamic range. As input it takes a regularly sampled density field. In a first step, the input field is Gaussian smoothed using a $\log$Filter filter that is applied over a set of scales $[R_0,R_1,...,R_N]$, with $R_n=2^{n/2}R_0$. It produces the logarithmic density field:
\begin{equation}
  \delta_+\,=\,\log (1+\delta (\mathbf{ x}))\,,
\end{equation}
The logarithmic density field of \nexus{} is better equipped to take account of the wide dynamic range of the nonlinear hierarchically evolved density field. The nonlinear field is highly non-Gaussian, with a large part of the volume having low-density values in combination with long high-density tails in the high-density cluster and filament regions. It translates into a nonlinear density field pdf that approaches a lognormal or skewed lognormal function. 

For each of the included scale-space scales, \nexus{} computes the eigenvalues of the Hessian matrix of the smoothed logarithmic density field. Using the Hessian eigenvalues of these, \nexus{} computes an environmental signature for each volume element that characterises how close this region is to an ideal knot, filament and wall. Then, for each point, the environmental signatures computed for each scale are combined to obtain a scale independent signature.

In the last step, physical criteria are used to determine a detection threshold. All points with signature values above the threshold are valid structures. For knots, the threshold is given by the requirement that most knot-regions should be virialized. For filaments and walls, the threshold is determined on the basis of the change in filament and wall mass as a function of signature. The peak of the mass variation with signature delineates the most prominent filamentary and wall features of the cosmic web.

The \nexus{} algorithm performs the environment detection by applying the above steps first to knots, then to filaments and finally to walls. Each volume element is assigned a single environment characteristic by requiring that filament regions cannot be knots and that walls regions cannot be either knots or filaments. The remaining regions are classified as voids. An example of this classification, showing the density field together with the corresponding \nexus{} signatures for nodes, filaments and walls, is presented in Figure \ref{fig:nexus}.


\bsp	
\label{lastpage}
\end{document}